\documentclass[12pt,conference, onecolumn]{IEEEtran}
\IEEEoverridecommandlockouts
\usepackage{epsfig, graphicx, extarrows, bm, titletoc, amsmath, amssymb, epsf, cite, subfigure, scalefnt, epstopdf, algorithm, algorithmic, fancyhdr, color}

    \def\by{{\mathbf{y}}}

   \def\bI{{\mathbf{I}}}

\def\b0{{\mathbf{0}}}

  \def\ibc{{\pmb{c}}} \def\ibd{{\pmb{d}}} 
    
   \def\ibn{{\pmb{n}}} 
\def\ibp{{\pmb{p}}}  \def\ibr{{\pmb{r}}} \def\ibs{{\pmb{s}}} 
  \def\ibw{{\pmb{w}}} \def\ibx{{\pmb{x}}} \def\iby{{\pmb{y}}}
\def\ibz{{\pmb{z}}}

\def\ibA{{\pmb{A}}}    
  \def\ibH{{\pmb{H}}} \def\ibI{{\pmb{I}}}

     \def\d4{\!\!\!\!}

\def\lp{\left(}     \def\rp{\right)}            \def\lS{ \left[ }
\def\rS{ \right] }

\def\mul{\! \times \!}  \def\-{\! - \!}  \def\+{\! + \!}  \def\={\! = \!}  \def\>{\! > \!}

\newtheorem{proposition}{\bf Proposition}
\newtheorem{assumption}{\bf Assumption}
\newtheorem{theorem}{\bf Theorem}
\newtheorem{lemma}{\bf Lemma}

\newtheorem{remark}{\bf Remark}

\newtheorem{property}{\bf Property}

\newcommand{\bef}{\begin{figure}}
\newcommand{\eef}{\end{figure}}
\newcommand{\beq}{\begin{eqnarray}}
\newcommand{\eeq}{\end{eqnarray}}

\newcommand{\qed}{\nobreak \ifvmode \relax \else
\ifdim\lastskip<1.5em \hskip-\lastskip \hskip1.5em plus0em
minus0.5em \fi \nobreak \vrule height0.5em width0.5em
depth0.25em\fi}

\graphicspath{{figures/}}

\newcommand{\BE}{\begin{equation}}
\newcommand{\EE}{\end{equation}}
\newcommand{\BS}{\begin{subequations}}
\newcommand{\ES}{\end{subequations}}
\newcommand{\revise}{\textcolor{black}}

\def\d4{\!\!\!\!}
\def\b0{{\mathbf{0}}}
\begin{document}
\lhead{}
\chead{}
\rhead{}
\lfoot{}
\cfoot{\thepage}
\rfoot{}
\renewcommand{\headrulewidth}{0pt}
\renewcommand{\footrulewidth}{0pt}
\pagestyle{fancy}

\title{Compressed Coding, AMP Based Decoding and Analog Spatial Coupling \\
\thanks{S. Liang, C. Liang and L. Ping are with Department of Electrical Engineering, City University of Hong Kong, Hong Kong SAR, China (e-mail: ssliang3-c@my.cityu.edu.hk, liangchulong@qq.com, eeliping@cityu.edu.hk). J. Ma is with the John A. Paulson School of Engineering and Applied Sciences, Harvard University, Cambridge, MA 02138, USA (e-mail: junjiema@seas.harvard.edu).} \thanks{This paper was presented in part at the 2016 9th International Symposium on Turbo Codes and Iterative Information Processing \cite{Liang16} and submitted in part to GLOBECOM 2020 \cite{Liang2020}.}}
\author{Shansuo~Liang, Chulong~Liang, Junjie~Ma, and~Li~Ping,~\IEEEmembership{Fellow,~IEEE } }
\maketitle
\begin{abstract}
This paper considers a compressed-coding scheme that combines compressed sensing with forward error control coding. Approximate message passing (AMP) is used to decode the message. Based on the state evolution analysis of AMP, we derive the performance limit of compressed-coding. We show that compressed-coding can approach Gaussian capacity at a very low compression ratio. Further, the results are extended to systems involving non-linear effects such as clipping. We show that the capacity approaching property can still be maintained when generalized AMP is used to decode the message.

To approach the capacity, a low-rate underlying code should be designed according to the curve matching principle, which is complicated in practice. Instead, analog spatial-coupling is used to avoid sophisticated low-rate code design. In the end, we study the coupled scheme in a multiuser environment, where analog spatial-coupling can be realized in a distributive way. The overall block length can be shared by many users, which reduces block length per-user.
\end{abstract}
\begin{IEEEkeywords}
Compressed sensing, forward error control coding, approximate message passing, state evolution, area theorem and analog spatial-coupling.
\end{IEEEkeywords}
\section{Introduction}

\revise{This paper is concerned with the following compressed-coding scheme:
\BE \label{x=Ac}
\ibx = \ibA \ibc,
\EE
where $\ibc \in \mathbb{R}^{N \mul 1}$ is an forward error control (FEC) coded and modulated sequence and $\ibA \in \mathbb{R}^{M \mul N}$ refers to a compression matrix \cite{Liang16,cLiang17}. We denote $\delta \triangleq M/N$ as the compression ratio of $\ibA$. For simplicity we consider the real case for \eqref{x=Ac} and the results can be directly extended to complex cases \cite{Ma19,liu2019capacity}.}

\revise{Sparse regression codes, introduced in \cite{Barron12,Joseph14,SRC19}, can also be represented by \eqref{x=Ac}. We will discuss this in Section \ref{Equivalence} in detail using the equivalence between position modulation (PM) and Hadamard coding. Approximate message passing (AMP), originally developed for compressed sensing, has been applied to decode sparse regression codes \cite{Donoho2009,Rush17,Barbier17AMP-decoder}. The performance of AMP can be analyzed using a state evolution (SE) technique \cite{SEBayati,Javanmard13}. It has been shown that detection algorithms based AMP and the related orthogonal AMP (OAMP) can potentially outperform the conventional Turbo-type detection algorithms in coded linear systems as \eqref{x=Ac} \cite{Ma19,liu2019capacity,Jeon15}.}

Spatial-coupling offers improved performance for Turbo and LDPC type codes \cite{Kudekar13,Mitchell15}. Most works on spatial-coupling are based on binary additions \cite{Kudekar13,Mitchell15,Ma15,Hou16}. Analog spatial-coupling over real or complex fields have been investigated for applications involving code-division multiple-access (CDMA) \cite{Takeuchi11}, data-coupling \cite{Truhachev13a} \revise{and compressed sensing \cite{Kudekar10,Krzakala12,Donoho2012}}. It is shown that data-coupling can approach Gaussian capacity at a asymptotically high signal-to-noise-ratio (SNR) \cite{Truhachev13a}. This is theoretically interesting, but the SNR range is outside the scope of most practical systems.

\revise{It has been shown that spatially-coupled sparse regression code (SC-SRC) is asymptotically capacity achieving \cite{Barbier17AMP-decoder,Barbier17GAMP,Rush20}. However, SC-SRC requires a very small $\delta$ for good performance at low-to-medium SNRs. For $\ibA$ with a fixed $M$, $N$ grows as $\delta$ decreases, which incurs an increase of memory for storing and decoding complexity. For this reason, most available simulation results of SC-SRC are for high SNR scenarios \cite{Barbier17AMP-decoder,Rush20,Hsieh20}.}

In this paper we study the scheme in (1) involving general FEC codes. We will show that combining some powerful techniques from signal processing and communications, namely compressed sensing, concatenated FEC coding, AMP based decoding and spatial-coupling, can offer significant performance gains. Our main findings are as follows.
\begin{itemize}
  \item We derive the performance limit of compressed-coding with AMP based decoding. Our basic assumption is that the SE for AMP remains accurate in the presence of an FEC decoder. Based on the area property of extrinsic information transfer charts, we show that compressed-coding can approach Gaussian capacity, even though the underlying coded sequence is non-Gaussian. This is consistent with the Gaussian distribution of the signals after compression.
  \item We further extend the results to systems involving non-linear effects such as clipping and quantization. We show that the near-optimal performance can be maintained when generalized AMP (GAMP) is used for decoding. This alleviates the problem of high peak-to-average-power ratio (PAPR) related to Gaussian signaling. Incidentally, sparse regression codes suffer from the same problem \cite{sLiang17}.
  \item The above capacity approaching property requires careful code optimization using the curve matching technique, which is rate specific and lacks flexibility \cite{Yue07,Divsalar11,Ping03}. We will introduce a spatially-coupled compressed-coding (SC-CC) scheme to circumvent this difficulty. \revise{Compared with SC-SRC, SC-CC can offer good performance over a wider SNR range. We will provide a graphic illustration that clearly explains this advantage of SC-CC over SC-SRC.}
  \item A code is said to be universal if it remains good after random puncturing. Such codes are useful in, e.g., type-II automatic repeat request (ARQ) applications \cite{Linshu04,Huang07,Zhang09}. The existing high rate coded modulation methods typically do not work well after heavy puncturing. We show that SC-CC is inherently universal and potentially capacity approaching at both low and high SNRs.
  \item We study SC-CC in a multiuser environment. A traditional view is that spatial-coupling will increase overall block length, which causes difficulty in applications with stringent latency requirements. Interestingly, in a multi-user system, analog spatial-coupling can be realized in a distributive way and the overall length can be shared by many users, which effectively reduces block length per-user. This offers an interesting new solution for future multi-user wireless communication systems.
\end{itemize}
In summary, the proposed SC-CC scheme offers practical solutions to some open challenges in coding techniques: (i) a simple method to approach the ultimate capacity of Gaussian signaling (beyond that of discrete signaling), (ii) a simple treatment of non-linear effects during transmission, (iii) a low-cost universal coding and decoding strategy and (iv) a multi-user scheme with short per-user block length and good performance. These claims are supported by extensive theoretical and numerical results.
\section{Compressed-Coding Scheme} \label{SecUncoupled}
\subsection{Compressed-Coding}

\begin{figure}[!htbp]
\centering
  \includegraphics[width=0.45\textwidth]{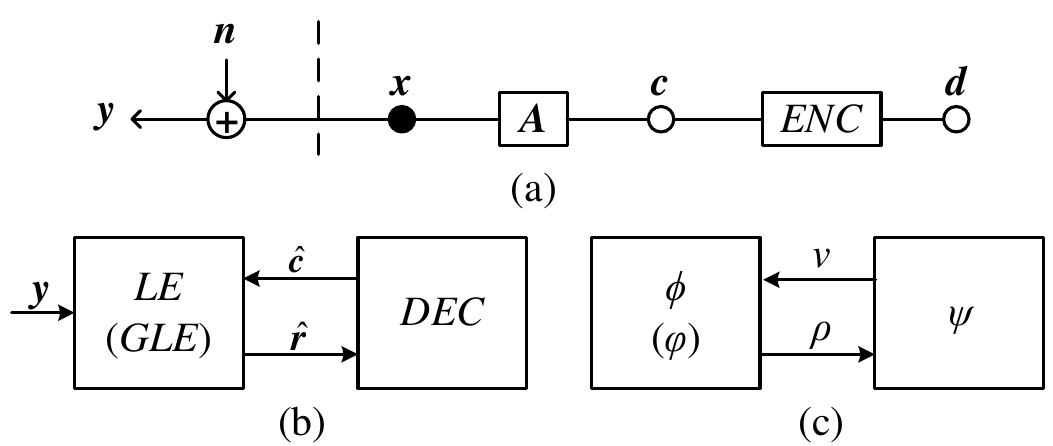}
  \vspace{-0.6cm}
  \caption{Graphic illustrations for (a) system model, (b) the receiver structures and (c) state evolution. ENC and DEC denote the encoder and decoder, respectively.}\vspace{-0.6cm}
  \label{system}
\end{figure}

\revise{Fig. \ref{system}(a) illustrates the system model for compressed-coding. A binary information sequence $\ibd \in \mathbb{B}^{J \mul 1}$ is encoded into $\ibc \in \mathbb{R}^{N \mul 1}$ based on an FEC code $\mathcal{C}$ and a constellation $\mathcal{S}$.} We assume that the entries of $\ibc$ are drawn from $\mathcal{S}\equiv \{ s_j \}$ with equal probabilities and $\mathrm{E}[s]=0$, $\mathrm{E}[|s|^2]=1$. Consider transmitting $\ibx=\ibA\ibc$ in an additive while Gaussian noise (AWGN) channel. The received signal $\iby \in \mathbb{R}^{M \mul 1}$ is given by 
\beq \label{AWGNCode}
\iby = \ibA \ibc + \ibn,
\eeq
where $\ibn \sim \mathcal{N}(\b0,\sigma^2\bI)$ contains AWGN samples. For theoretical analysis, we assume that the entries of $\ibA \in \mathbb{R}^{M \mul N}$ are independent and identically distributed (i.i.d.) as $A_{m,n} \sim \mathcal{N}(0,1/N)$\footnote{\revise{Note that $\ibA$ is row-normalized to unit in this paper, while $\ibA$ is column-normalized to unit in the original AMP algorithm in \cite{Donoho2009}. We adopt row-normalization here to ensure that, at the fixed symbol power of $\ibc$, the symbol power of $\ibx$ does not change with $\delta$. This makes the capacity expression in \eqref{Eqn:phi}--\eqref{Eqlemma1} relatively simpler.}}. \revise{ The information rates of $\ibc$ and $\ibA \ibc$ are defined as $R_{\text{C}} \triangleq J/N$ and $R_{\text{AC}} \triangleq J/M$ bits per channel use (bpcu), respectively. Recall that the compression ratio of $\ibA$ is $\delta = M/N$, thus we have $R_{\text{AC}} = R_{\text{C}} / \delta$.} When $R_{\text{C}}$ is fixed, $R_{\text{AC}}$ can be adjusted by choosing different $\delta$. The task at the receiver is to recover $\ibd$ from $\iby$ with known $\ibA$.
\revise{\subsection{Connection to Sparse-Regression Codes}\label{Equivalence}
Sparse regression code can be represented by \eqref{x=Ac} with $\ibc$ segmented as $\ibc^T = [\ibc_1^T,...,\ibc_l^T,...,\ibc_L^T ]$, where each $\ibc_l \in \mathbb{R}^{B \mul 1}$ contains exact one entry of ``1'' and $B-1$ ``0''. Such PM is equivalent to Hadamard coding. Too see this, let $\ibH \in \mathbb{R}^{B \mul B}$ be a Hadamard matrix over $\{-1, +1\}$ \cite{Leung06}. The columns of $\ibH$ form a Hadamard code. For $\ibc_l$ defined above, $\tilde{\ibc}_l = \ibH \ibc_l$ is a Hadamard codeword. Define a block diagonal matrix $\widetilde{\ibH} \triangleq \mathrm{diag}[\ibH,\ibH,...,\ibH] \in \mathbb{R}^{LB \mul LB}$. Then $\tilde{\ibc} \equiv \widetilde{\ibH} \ibc$ can be segmented as $\tilde{\ibc}^T = [\tilde{\ibc}_1^T,...,\tilde{\ibc}_l^T,...,\tilde{\ibc}_L^T ]$, with each $\tilde{\ibc}_l$ being a Hadamard codeword. Due to the orthogonality of Hadamard matrices ($\widetilde{\ibH}^T\widetilde{\ibH}=B \cdot \ibI$), we have
\BE
\ibx = \ibA \ibc = B^{-1} \ibA \widetilde{\ibH}^T\widetilde{\ibH} \ibc =  \widetilde{\ibA} \tilde{\ibc},
\EE
where $\widetilde{\ibA} \triangleq B^{-1} \ibA \widetilde{\ibH}^T$. Therefore, $\ibx$ can be generated using either PM ($\ibx = \ibA \ibc$) or Hadamard coding ($\ibx = \widetilde{\ibA} \tilde{\ibc}$). Assume that the entries of $\ibA$ are i.i.d Gaussian. Due to the orthogonality of $\widetilde{\ibH}$, the entries of $\widetilde{\ibA}$ are also i.i.d Gaussian. Statistically, $\ibA$ and $\widetilde{\ibA}$ are equivalent. This clearly shows the equivalence between sparse regression code and compressed-coding using a Hadamard code. Numerical results for such equivalence will be given in Fig. \ref{SimulationV1}.}
\subsection{AMP-based Decoding} \label{SectionSE}
Initializing from $\ibc^1 = \b0$ and $\ibr^1_{\mathrm{Onsager}}=\b0$, \revise{ the AMP algorithm alternates between a linear estimator (LE) and an nonlinear estimator (NLE) as \footnote{\revise{In fact, AMP can be extended to a more general case, where $A_{m,n} \sim \mathcal{N}(0,\sigma_a^2/M)$. In this case, we can rewrite the system to $\iby'=\sigma_a^{-1}\iby = \ibA'\ibc+\ibn'$, where $A'_{m,n} \sim \mathcal{N}(0,1/M)$ and $\ibn' \sim \mathcal{N}(\b0,\sigma_a^{-2}\sigma^2\bI)$. Then, the original AMP algorithm and SE in \cite{Donoho2009} can be applied by replacing $\iby$, $\ibA$, $\sigma^2$ with $\iby'$, $\ibA'$, $\sigma_a^{-2}\sigma^2$, respectively. For example, for the column normalization considered in this paper, we set $\sigma_a^{2}=\delta$ to make the results of this paper valid.}} \cite{Donoho2009,Barbier17AMP-decoder,Rush17}
\BS \label{AMP}
\begin{eqnarray} \label{AMPLE}
  \mathrm{LE:}&&\d4\d4 \quad \ibr^t =  \hat{\ibc}^t + \delta \ibA^\textmd{T} (\iby-\ibA \hat{\ibc}^t ) + \ibr^t_{\mathrm{Onsager}},\\
  \mathrm{NLE:}&&\d4\d4 \quad \hat{\ibc}^{t+1} = \eta (\ibr^t), \label{APPDEC}
\end{eqnarray}
\ES
where $\eta$ is a denoising function of $\ibr^t$. In \eqref{AMPLE}, $\ibr^t_{\mathrm{Onsager}}$ is an ``Onsager'' term defined as
\beq \label{Onsager}
\ibr^t_{\mathrm{Onsager}} \triangleq \frac{1}{\delta}\langle \eta' (\ibr^{t-1})  \rangle \left( \ibr^{t-1} - \ibc^{t-1} \right),
\eeq
where $\langle \cdot \rangle$ denotes the average of the inputs.} The final estimate is given by the hard decision based on $\hat{\ibc}^{T+1}$, where $T$ is the maximum number of iterations. LE is used to handle the linear observation constraint $\iby = \ibA \ibc + \ibn$ while NLE is used to explore the prior information of $\ibc$. The Onsager term is used to regulate correlation among messages during iterative processing.

Fig. \ref{system}(b) illustrates an iterative receiver for \eqref{AWGNCode} based on AMP. The denoising function $\eta$ in \eqref{APPDEC} is given by an \textit{a posteriori} probability (APP) decoder for the underlying FEC code $\mathcal{C}$. Its input $\ibr^t$ generated in \eqref{AMPLE} is treated as a noisy observation of $\ibc$ using the following model \cite{Barbier17AMP-decoder,Rush17}
\BE \label{AWGN-x}
\ibr^t = \ibc + (\rho^t)^{-1/2} \cdot \ibw,
\EE
where $\ibw \sim \mathcal{N}(\b0,\bI)$ is independent of $\ibc$ and $\rho^t$ is the equivalent channel SNR. \revise{The output of $\eta$ is the APP mean of $\ibc$ based on $\ibr^t$ in \eqref{AWGN-x} as
\BE \label{eta}
\hat{\ibc}^t = \eta(\ibr^t) \triangleq \mathrm{E}[\ibc|\ibr^t,\mathcal{S},\mathcal{C}].
\EE
In practice, we compute \eqref{eta} as follows. We feed $\ibr^t$ into a soft-output decoder that generates the APP log-likelihood ratio (LLR) for each coded bit. Such APP LLRs are then used to generate the APP mean $\hat{\ibc}^t$ (see \cite[Section IV-A]{Ping10}). The corresponding mean squared-error (MSE) is denoted as
\BE \label{psi-opt}
\psi(\rho^t) = \frac {1} {N} \cdot \textrm{E}\left[\|\ibc-\textrm{E}(\ibc|\ibr^t,\mathcal{S},\mathcal{C})\|^2 \right],
\EE
where the expectation is over the distribution of $\ibw$ in \eqref{AWGN-x} and the underlying code $\mathcal{C}$ (including modulation over $\mathcal{S}$).}

\subsection{Evolution Analysis} \label{SectionSE}
\revise{Define the large system limit as $M,N \to \infty$ with a fixed $\delta \in (0, \infty) $.} The MSE performance of AMP in the large system limit can be tracked by a scalar SE recursion \cite{SEBayati,Donoho2009}. For the $t$-th iteration, let $v^t$ be the \textit{a priori} variance at the input of LE and $\rho^t$ the \textit{a priori} SNR at the input of the FEC decoder. Initializing with $v^1=1$, the SE recursion at the $t$-th iteration is given by \cite{SEBayati}
\BE \label{SE-AMP}
\rho^t = \phi(v^t) \quad \textrm{and} \quad v^{t+1} = \psi(\rho^t),
\EE
where $\phi(v^t)$ gives the SNR at LE output and $\psi(\rho^t)$ gives the variance at decoder output. Fig. \ref{system}(c) illustrates the SE recursion.

\revise{The accuracy of SE was proved for AMP and GAMP under the assumption that $\eta$ is ``separable'' \cite{SEBayati,Javanmard13}. In general, an FEC decoder cannot be regarded as ``separable''. The discussions in this paper are based on Assumption \ref{AssumpSEAMP} below. Simulation results will be provided to support this assumption.
\begin{assumption} \label{AssumpSEAMP}
SE is accurate for both AMP and GAMP algorithms involving an FEC decoder in the large system limit including a fixed $\delta$ that is arbitrarily close to zero.
\end{assumption}}

\begin{figure*}[!t]
\centering
\subfigure[]{
\begin{minipage}[t]{0.375\textwidth}
\centering \label{Transfer1-V1}
\includegraphics[width=1\textwidth]{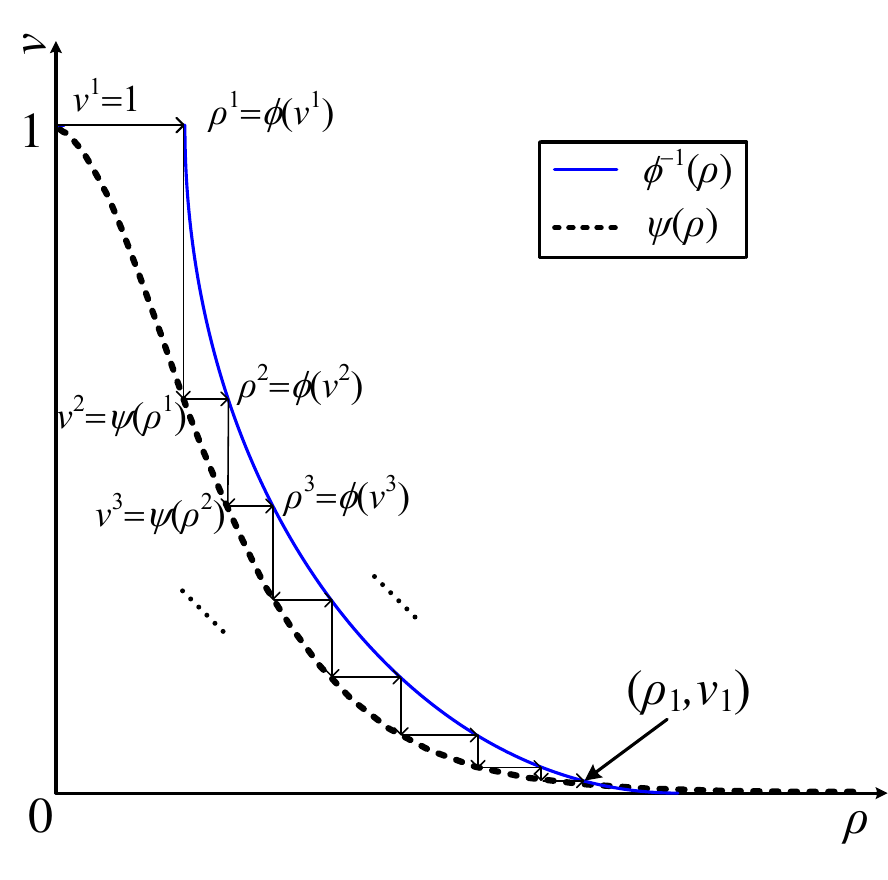}
\end{minipage} }
\subfigure[]{
\begin{minipage}[t]{0.375\textwidth}
\centering \label{Transfer_BPSK}
\includegraphics[width=1\textwidth]{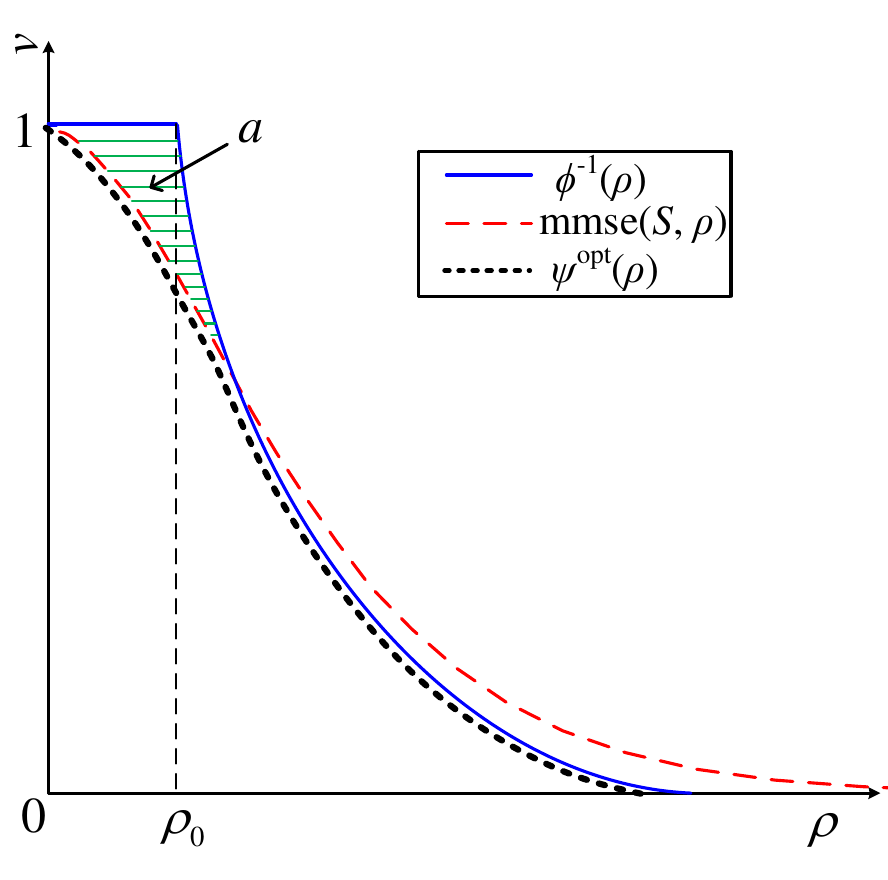}
\end{minipage}}
\caption{Graphic illustrations for (a) an SE  recursion with a general decoding function $\psi$ and (b) the optimal $\psi^{opt}$ satisfying the matching condition. The functions $\phi$ in two figures are the same.} \label{CurveMatch} \vspace{-0cm} \label{Transfer1-V2}
\end{figure*}

\revise{Fig. \ref{Transfer1-V1} illustrated the SE recursion for a general $\psi$. Starting from $v^1=1$, the zigzag curve between $\phi$ and $\psi$ illustrates the iterative recovery trajectory as $v^{\infty} = \psi(\phi( \cdots \psi(\phi(v^1=1))))$. The fixed point of SE is given by the first intersection of $\phi$ and $\psi$ shown as $(\rho_1,v_1)$. The final MSE performance of the AMP algorithm is given by the variance of the fixed point, i.e., $v_1$.}
\begin{property} \label{Prop1}
Error-free decoding is achieved when $v_1 \rightarrow 0$, which requires $\psi(\rho) \leq \phi^{-1}(\rho)$ for $\rho \in [0,\infty)$ with $v=\phi^{-1}(\rho)$ being the inverse function of $\rho=\phi(v)$.
\end{property}

\subsection{Area Property} \label{SectProperty}
According to \cite{Donoho2009}, $\phi(v)$ in (\ref{SE-AMP}) is given by
\BE \label{Eqn:phi}
\rho=\phi(v)=\frac{\delta}{v+\sigma^2}, \quad 0 \leq v \leq 1.
\EE
The capacity of a real-valued AWGN channel with Gaussian signaling is given by
\BE \label{C-AWGN}
C_{\text{G}}=0.5\log(1+1/\sigma^2).
\EE
From \eqref{Eqn:phi}--\eqref{C-AWGN}, we have the following lemma:
\begin{lemma} \label{lemma1}
\BE \label{Eqlemma1}
\frac{1}{2}\int_0^1 {\phi \left( v \right){\rm{d}}v}  = \delta  \cdot C_{{\rm{G}}}.
\EE
\end{lemma}
The left hand side of \eqref{Eqlemma1} can be interpreted as the area under $\phi$ scaled by 0.5 (for a real-valued channel). Lemma \ref{lemma1} bridges the area under $\phi$ and the AWGN channel capacity.


Assume that $\eta$ for decoding $\ibc$ in \eqref{eta} is Bayes-optimal. Then $\psi$ in \eqref{psi-opt} gives the minimum MSE for decoding $\ibc$. The lemma below follows \cite[Corollary 1]{Bhattad2007}:
\begin{lemma} \label{lemma2}
\BE \label{Eqn:lemma2}
\frac{1}{2}\int_0^\infty  {\psi \left( \rho  \right){\rm{d}}} \rho  = R_{\text{C}}.
\EE
\end{lemma}

\subsection{Achievable Rate} \label{AchievableRate}
Define the inverse function of $\phi$ in \eqref{Eqn:phi} as
\beq \label{Defphi}
v=\phi^{-1}(\rho)=\left\{
       \begin{array}{ll}
         1, & \hbox{$ 0 \leq \rho < \rho_0 $,} \\
         \frac{\delta}{\rho} - \sigma^2, & \hbox{$\rho \geq \rho_0$,}
       \end{array}
     \right.
\eeq
where $\rho_0 \triangleq \frac {\delta} {1+\sigma^2}$ is shown in Fig. \ref{Transfer_BPSK}.

\revise{Let $\textrm{mmse}(\mathcal{S},\rho)$ be the MMSE for the symbol-by-symbol detection (without the coding constraint $\mathcal{C}$):
\BE \label{mmse-opt}
\textrm{mmse}(\mathcal{S},\rho) = \textrm{E}\left[|c-\textrm{E}(c|c+\rho^{-1/2} w,\mathcal{S})|^2 \right],
\EE
where $w \sim \mathcal{N}(0,1)$ is independent of $c$ and the expectation is taken over $\mathcal{S}$. The symbol-by-symbol estimation in \eqref{mmse-opt} cannot outperform the one in \eqref{psi-opt} since the latter considers the coding constraint.} Thus, we have $\psi(\rho) \leq \textrm{mmse}(\mathcal{S},\rho)$ for $\rho \in [0,\infty)$. Combining with Property \ref{Prop1} yields
\beq \label{PhiUpper}
\psi(\rho) \leq \mathrm{min} \{ \phi^{-1}(\rho),\textrm{mmse}(\mathcal{S},\rho) \}, \quad \rho \in [0,\infty).
\eeq

According to Lemma \ref{lemma2}, the achievable rate of $\ibc$ can be maximized by maximizing the area under $\psi$. From \eqref{PhiUpper}, the maximal area is achieved when
\beq \label{MatchingS}
\psi^{opt}(\rho) = \mathrm{min} \{ \phi^{-1}(\rho),\textrm{mmse}(\mathcal{S},\rho) \}, \quad \rho \in [0,\infty).
\eeq
Based on \eqref{MatchingS}, we define
\beq \label{AreaS}
a \triangleq \int_0^\infty \left( \phi^{-1}(\rho) -  \psi^{opt}(\rho)  \right) {\rm{d}} \rho.
\eeq
\revise{Fig. \ref{Transfer_BPSK} shows examples of $\psi^{opt}(\rho)$ and $a$ that are illustrated by the dot line and the shadowed area, respectively.} According to Lemmas \ref{lemma1}--\ref{lemma2}, we have
\BE \label{as2}
a = \int_0^1 \phi(v) {\rm{d}} v -  \int_0^\infty \psi^{opt}(\rho) {\rm{d}} \rho = 2 \delta C_{\text{G}} - 2 R_{\text{C}}.
\EE
The overall rate after the compression matrix is given by
\beq \label{R_CS_S}
R_{\text{AC}} = R_{\text{C}}/\delta = C_{\text{G}} - a / 2\delta.
\eeq
Eq. \eqref{R_CS_S} shows that the rate of compressed-coding is away from the AWGN capacity by a gap of $a / 2\delta$. In the next subsection, we show that this gap vanishes as $\delta, R_{\text{C}}\to0$ with $R_{\text{AC}} = R_{\text{C}} / \delta$ fixed.
\subsection{Approaching Capacity} \label{RateGap}
We first consider  BPSK for $\mathcal{S}$. Afterwards, we extend the results to more general cases.

For BPSK, the entries of $\ibc$ are drawn from $\mathcal{B} = \{ +1,-1 \}$ with equal probabilities. From \eqref{MatchingS}, we have
\beq \label{MatchingBPSK}
\psi_{\mathcal{B}}^{opt}(\rho) = \mathrm{min} \{ \phi^{-1}(\rho),\textrm{mmse}(\mathcal{B},\rho) \}, \quad \rho \in [0,\infty),
\eeq
where $\textrm{mmse}(\mathcal{B},\rho)$ is the MMSE for detecting BPSK from an AWGN channel \cite{Guo2011}:
\beq \label{Eqn:MMSE_b}
\textrm{mmse}(\mathcal{B},\rho) = 1 - \int_{ - \infty }^{ + \infty } {\frac{{e^{ - x^2 /2} }}{{\sqrt {2\pi } }}} \tanh \left( {\rho  - \sqrt \rho  x} \right){\textrm{d}}x.
\eeq
Following \eqref{AreaS}, we define
\BE \label{as_BPSK}
a_{\mathcal{B}} \triangleq \int_0^\infty \left( \phi^{-1}(\rho) -  \psi_{\mathcal{B}}^{opt}(\rho)  \right) {\rm{d}} \rho.
\EE
The corresponding rate is given by
\beq \label{R_CS_ori}
R_{\text{AC}} = C_{\text{G}} - a_{\mathcal{B}} / 2\delta.
\eeq
\begin{theorem}\label{delta/a2zero1}
Assume that the matching condition \eqref{MatchingBPSK} holds for BPSK. Then, $R_{\text{AC}} \to C_{\text{G}}$ when $\delta, R_{\text{C}}\to0$ with $R_{\text{C}}/\delta$ fixed.
\end{theorem}
\begin{IEEEproof}
See Appendix \ref{proofdelta/a2zero1}.
\end{IEEEproof}


Next, we consider a commonly used symmetric constellation $\mathcal{S}_C$ such that if $s \in \mathcal{S}_C$ then $-s \in \mathcal{S}_C$. Such a $\mathcal{S}_C$ includes quadrature phase shift keying (QPSK) and quadrature amplitude modulation (QAM) as special examples. For such $\mathcal{S}_C$, similar to \eqref{R_CS_ori}, we can show that
\BE \label{R_CS_Sym}
R_{\text{AC}} = C_{\text{G}} - a_{\mathcal{S}_C} / 2\delta,
\EE
where $a_{\mathcal{S}_C} = \int_0^\infty \left( \phi^{-1}(\rho) -  \psi_{\mathcal{S}_C}^{opt}(\rho)  \right) {\rm{d}} \rho$ with the following matching condition
\BE \label{MatchingSC}
\psi_{\mathcal{S}_C}^{opt}(\rho) = \mathrm{min} \{ \phi^{-1}(\rho),\textrm{mmse}(\mathcal{S}_C,\rho) \}, \quad \rho \in [0,\infty).
\EE
\begin{theorem}\label{Theom2}
Assume that the matching condition \eqref{MatchingSC} holds for $\mathcal{S}_C$. Then, $R_{\text{AC}} \to C_{\text{G}}$ when $\delta, R_{\text{C}}\to0$ with $R_{\text{C}}/\delta$ fixed.
\end{theorem}
\begin{IEEEproof}
Due to the symmetry, we can treat $\mathcal{S}_C$ as the sum of multiple BPSK constellations multiplied by proper scalings.
Recall the assumption that constellation points are drawn from $\mathcal{S}_C$ with equal probability. According to \cite[Proposition 14]{Guo2011}, we have $\textrm{mmse}(\mathcal{S}_C,\rho) \geqslant \textrm{mmse}(\mathcal{B},\rho)$ for $\rho \in [0,\infty)$.

Comparing \eqref{MatchingBPSK} and \eqref{MatchingSC} yields $\psi_{\mathcal{S}_C}^{opt}(\rho) \geqslant \psi_{\mathcal{B}}^{opt}(\rho)$ for $\rho \in [0,\infty)$. Furthermore, we have $a_{\mathcal{S}_C} \leq a_{\mathcal{B}}$ according to $a$ defined in \eqref{AreaS}. Recall Theorem \ref{delta/a2zero1} that $a_{\mathcal{B}} / 2\delta \to0$ when $\delta, R_{\text{C}}\to0$ with $R_{\text{C}}/\delta$ fixed. Since $a_{\mathcal{S}_C} \leq a_{\mathcal{B}}$, $a_{\mathcal{S}_C} / 2\delta \to0$ also holds. From \eqref{R_CS_Sym}, we have $R_{\text{AC}} \to C_{\text{G}}$ and complete the proof.
\end{IEEEproof}

\begin{remark} \label{remark1}
Theorem \ref{Theom2} shows that $R_{AC}$ of compressed-coding with $\mathcal{S}_C$ can approach $C_G$ under two sufficient conditions: (a) the matching condition \eqref{MatchingSC} holds and (b) both $R_C$ and $\delta$ $\to 0$. These conditions require an underlying low-rate FEC code $\mathcal{C}$ that meets the matching condition \eqref{MatchingSC}. In practice, it is a highly complicated task to design such a low-rate code (see \cite{Yue07,Divsalar11,Ping03} for details). There is another difficulty. Due to the Gaussian distribution of $\ibx$, the compressed-coding scheme suffers from a high PAPR problem.

In what follows, we will address these two difficulties in the next two sections separately.
\end{remark}

\section{Compressed-Coding with Clipping} \label{SecGAMP}
The aforementioned results are for the linear system $\iby = \ibA \ibc + \ibn$. In this section, we extend the results to a more general system modeled below:
\BE \label{Eqn:SysClip}
\iby = f(\ibx)+\ibn = f(\ibA \ibc)+\ibn,
\EE
where $f$ is a symbol-by-symbol function. This generalized scheme arises in various practical applications. An example is the clipping function for alleviating the high PAPR problem mentioned at the end of Section \ref{SecUncoupled} that is given by
\BE \label{Eqnclip}
f(x) \triangleq \left\{
       \begin{array}{ll}
         Z, & \hbox{$x > Z$,} \\
         x, & \hbox{$-Z \leq x \leq Z$,} \\
         -Z, & \hbox{$x < -Z$,}
       \end{array}
     \right.
\EE
where $Z>0$ is the clipping threshold.

Alternatively, consider a slightly different system model:
\BE \label{Eqnquan}
\iby = f(\ibA \ibc + \ibn ).
\EE
As an example, $f$ in \eqref{Eqnquan} may represent      the quantization effect of low-resolution analog-to-digital conversion on the received signal. In this section, we first focus on \eqref{Eqn:SysClip}. The treatment for \eqref{Eqnquan} is discussed in Appendix \ref{theo4extension}.
\subsection{GAMP and State Evolution}
At the receiver side, GAMP can be used to recover $\ibc$ from $\iby$ in \eqref{Eqn:SysClip} involving nonlinearity. Similar to AMP, the MSE performance of GAMP can be characterized by a SE recursion \cite{GAMP,Javanmard13}. We now briefly outline the GAMP algorithm and the corresponding SE recursion. Based on the SE, we analyze the achievable rate of GAMP for \eqref{Eqn:SysClip} following the procedure in Section \ref{SecUncoupled}.

When $M,N \rightarrow \infty$ with $\delta = M/N$ fixed, we have $\| \ibA\|^2_F / M  \rightarrow 1$ and $\| \ibA\|^2_F / N \rightarrow \delta$ since $A_{m,n} \sim \mathcal{N}(0,1/N)$ (see \eqref{AWGNCode}). Initializing $\hat{\ibc}^1=\b0$, $\ibs^0=\b0$ and $v^1=1$, the GAMP algorithm in \cite{GAMP} can be summarized by the following iteration between a generalized LE (GLE) and a NLE:
\BS \label{GAMP-alg}
\begin{eqnarray}
\mathrm{GLE:}&&\d4  \hat{\ibp}^t = \ibA\hat{\ibc}^t - v^t \ibs^{t-1}, \\
&&\d4 \ibs^t = \frac {g(\hat{\ibp}^t,\iby) -\hat{\ibp}^t} {v^t}, \label{GAMP-MMSE} \\ \label{GAMP-rt}
&&\d4 \ibr^t = \hat{\ibc}^{t} + \frac{v^t / \delta}{1 - \langle \partial g(\hat{\ibp}^t,\iby) / \partial \hat{\ibp}^t \rangle}\ibA^\textmd{T} \ibs^t, \\
\mathrm{NLE:}&&\d4 \hat{\ibc}^{t+1} = \eta (\ibr^t). \label{GAMP-eta}
\end{eqnarray}
\ES
For GLE, $g(\hat{\ibp}^t,\iby)$ in \eqref{GAMP-MMSE} performs the MMSE estimation of $\ibx$ based on its prior $\hat{\ibp}^t$ and the observation $\iby$ in \eqref{Eqn:SysClip} as
\beq \label{Eqn:MMSE-clip}
g(\hat{p}_m^t,y_m) =\textrm{E}[x_m|\hat{p}_m^t,y_m], \quad m = 1,2,...,M,
\eeq
where $p(\ibx|\hat{\ibp}^t) = \mathcal{N}(\hat{\ibp}^t,v^t \bI)$ at the $t$-th iteration. Similar to \eqref{AWGN-x}, $\ibr^t$ in \eqref{GAMP-rt} can be treated as an AWGN observation of $\ibc$. For NLE, $\eta$ is given by an FEC decoder as illustrated in Fig. \ref{system}(b).

The MSE performance of GAMP in \eqref{GAMP-alg} can be tracked by an SE recursion \cite{GAMP,Javanmard13}. With abuse of notation, let $v^t$ be the \textit{a priori} variance at the input of GLE and $\rho^t$ the \textit{a priori} SNR at the input of the decoder as shown in Fig. \ref{system}(c). Initializing with $v^1=1$, the SE recursion for \eqref{GAMP-alg} is given by \cite{GAMP,SEBayati}
\BE \label{SE-GAMP}
\rho^t = \varphi(v^t) \quad \textrm{and} \quad v^{t+1} = \psi(\rho^t),
\EE
where $\psi$ is the same as \eqref{psi-opt}, and $\varphi$ is given as follows.

Let $p(x,\hat{p})$ be a joint Gaussian distribution as \vspace{-0.3cm}
\BE \label{C_z_phat}
\left [x, \hat{p} \right]^T \sim \mathcal{N}(\b0,\mathbf{\Sigma}),
\EE
where \vspace{-0.3cm}
\BE \label{Cov_z_phat}
\mathbf{\Sigma} =
\left [
\begin{array}{cc}
1   &   1-v\\
1-v & 1-v
\end{array}
\right].
\EE
Denote $p(y|x)$ as the likelihood of \eqref{Eqn:SysClip}. According to \cite{GAMP}, $\varphi$ in \eqref{SE-GAMP} is given by
\beq \label{GAMPphi}
\rho = \varphi(v) = \frac {\delta} {v} \left( 1- \frac {\textrm{mmse}(x|\hat{p},y)} {v} \right),
\eeq
with \vspace{-0.3cm}
\beq \label{MMSE-clip}
\textrm{mmse}(x|\hat{p},y) \equiv \textrm{E}[|x-\textrm{E}(x|\hat{p},y)|^2],
\eeq
where the expectation is over $p(\hat{p},x,y)$. The following result follows \cite[Claim 1]{GAMP}:
\begin{property} \label{propMarkov}
	In \eqref{C_z_phat}--\eqref{MMSE-clip}, $\hat{p} \rightarrow x \rightarrow y$ forms a Markov chain. Thus, $p(\hat{p},x,y)=p(y|x)p(x,\hat{p})$, where $p(y|x)$ and $p(x,\hat{p})$ are given by \eqref{Eqn:SysClip} and \eqref{C_z_phat}, respectively.
\end{property}
\subsection{Area Property} \label{area-GAMP}
The area properties derived below will be useful for achievable rate analyses in the next subsection.

Since $\psi$ is still given by \eqref{psi-opt}, Lemma \ref{lemma2} applies to $\psi$ in \eqref{SE-GAMP}. Theorem \ref{Theo1} below is presented first to underpin the area property of $\varphi$ in Theorem \ref{Theo3}.

Consider three random variables $S_1,S_2,S_3$ that form a Markov chain $S_1 \to S_2 \to S_3$. Assume that $S_1$ and $S_2$ are jointly distributed as
\BE \label{joint_z_phat}
\left [ S_1, S_2 \right]^T \sim \mathcal{N}(\b0,\mathbf{\Sigma}),
\EE
where $\mathbf{\Sigma}$ is given by \eqref{Cov_z_phat}. $S_2 \to S_3$ is characterized by a likelihood as $p(S_3|S_2)$. Denote the mutual information between $S_1$ and $S_3$ as $I(S_1;S_3)$ and define
\BE
\mathrm{mmse}(S_2|S_1,S_3) \triangleq \textrm{E}[|S_2-\textrm{E}(S_2|S_1,S_3)|^2],
\EE
where the expectation is over $p(S_1,S_2,S_3)$.

\begin{theorem} \label{Theo1}
\BE \label{EqnTheorem}
- \frac {\partial} {\partial v} I(S_1;S_3) = \frac {1} {2v} \lp 1- \frac  {\mathrm{mmse}(S_2|S_1,S_3)} {v} \rp.
\EE
\end{theorem}
\begin{IEEEproof}
	See Appendix \ref{SectProof}.
\end{IEEEproof}

Denote $I(y;x)$ as the mutual information between $y$ and $x$ in \eqref{Eqn:SysClip}, where $x \sim \mathcal{N}(0,1)$. Theorem \ref{Theo3} below establishes an area property for $\varphi$.
\begin{theorem} \label{Theo3}
\beq \label{AreaPhiG}
\frac {1} {2} \int_0^1 \varphi(v) dv= \delta \cdot I(y;x).
\eeq
\end{theorem}
\begin{IEEEproof}
According to Property \ref{propMarkov}, applying Theorem \ref{Theo1} to the Markov chain $\hat{p} \rightarrow x \rightarrow y$ yields
\BE
- \frac {\partial} {\partial v} I(\hat{p};y) = \frac {1} {2v} \lp 1- \frac  {\mathrm{mmse}(x|\hat{p},y)} {v} \rp.
\EE
Taking integrations of the above equation yields
\BS \label{integration}
\begin{eqnarray}
\frac {1} {2} \int_0^1 \frac {1} {v} \lp 1- \frac  {\mathrm{mmse}(x|\hat{p},y)} {v} \rp dv =&&\d4\d4 \int_0^1 - \frac {\partial} {\partial v} I(y;\hat{p}) dv \\
=&&\d4\d4 \lS I(y;\hat{p}) \rS_{v=0} - \lS I(y;\hat{p})\rS_{v=1} \\
\overset{(a)}{=}&&\d4\d4 I(y;\hat{p}=x) - I(y;\hat{p}=0) \\
=&&\d4\d4 I(y;x),
\end{eqnarray}
\ES
where $(a)$ is due to $p(\hat{p},x)$ in \eqref{C_z_phat}--\eqref{Cov_z_phat}. Comparing \eqref{integration} and \eqref{GAMPphi}, it is clear to get \eqref{AreaPhiG}.
\end{IEEEproof}
\begin{remark}
Beyond \eqref{Eqn:SysClip}, Theorem \ref{Theo3} holds for general systems that can be modeled by $p(\iby|\ibx)=\prod_{m=1}^M p(y_m|x_m)$, since the underlying proof of Theorem \ref{Theo1} does not specify $p(S_3|S_2)$. This can be seen from the proof of Theorem \ref{Theo3}, where $p(S_3|S_2)$ characterizes the relationship between $\ibx$ and $\iby$ in \eqref{Eqn:SysClip}.
\end{remark}
\subsection{Achievable Rate} \label{Rate-GAMP}
With the area properties of $\psi$ and $\varphi$ obtained in Lemma \ref{lemma2} and Theorem \ref{Theo3}, respectively, we now evaluate the achievable rate of the generalized compressed-coding scheme according to the curve matching principle. The property below follows \cite[Lemma 4.2]{Barbier17GAMP}.
\begin{property} \label{Property1}
For $v \in [0,1]$, $\varphi(v)$ is positive and decreasing with $v$.
\end{property}

\begin{figure}[!t]
\centering
  \includegraphics[width=0.3\textwidth]{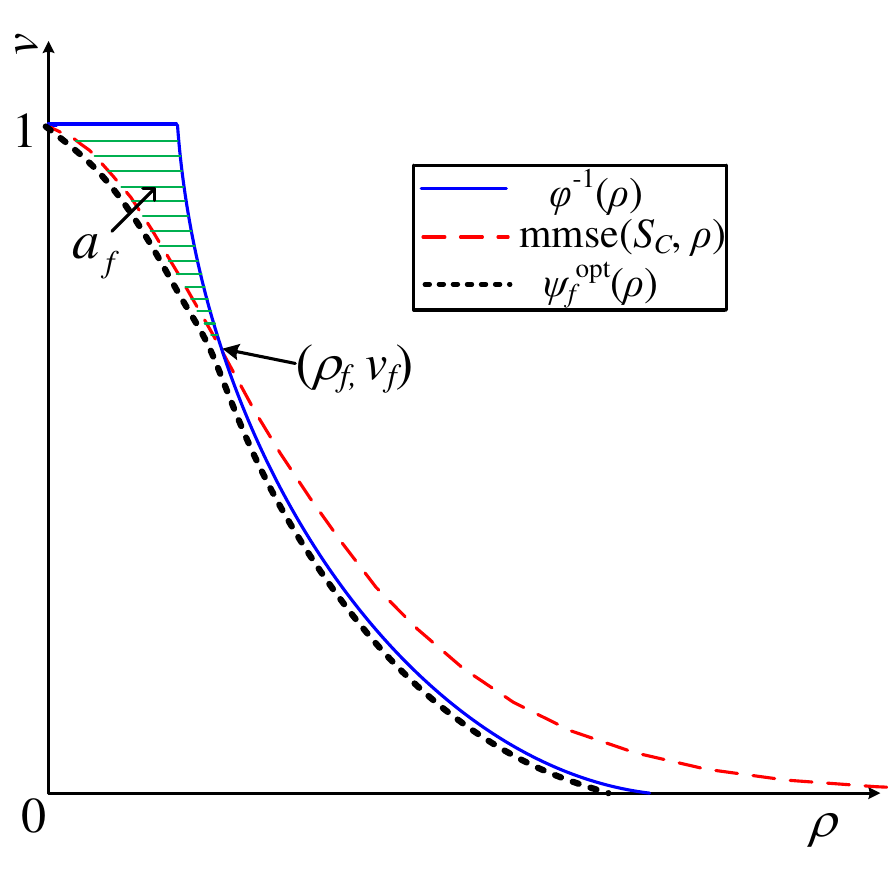}
  \vspace{-0.5cm}
  \caption{Examples of transfer functions for GAMP.} \vspace{-0.5cm}
  \label{Transfer_GAMP}
\end{figure}

From Property \ref{Property1}, $\rho = \varphi(v)$ is a one-to-one mapping from $v \in [0,1]$. Following the definition of $\phi^{-1}$ in \eqref{Defphi}, we define the inverse function of $\rho = \varphi(v)$ as $v \equiv \varphi^{-1}(\rho)$ that is shown by the solid line in Fig. \ref{Transfer_GAMP}.

Assume a symmetrical constellation $\mathcal{S}_C$ for $\ibc$. Similar to \eqref{PhiUpper}, $\psi$ in GAMP is upper bounded by
\beq \label{GAMP-psi-less}
\psi_f(\rho) \leq \mathrm{min} \{ \varphi^{-1}(\rho),\textrm{mmse}(\mathcal{S}_C,\rho) \}, \quad \rho \in [0,\infty).
\eeq
According to Lemma \ref{lemma2}, the achievable rate of $\psi$ can be maximized when the equality holds in \eqref{GAMP-psi-less}, i.e.,
\beq \label{GAMP-psi-opt}
\psi_f^{opt}(\rho) = \mathrm{min} \{ \varphi^{-1}(\rho),\textrm{mmse}(\mathcal{S}_C,\rho) \}, \quad \rho \in [0,\infty).
\eeq
Based on \eqref{GAMP-psi-opt}, define
\BE \label{areaf}
a_f \triangleq \int_0^{\infty} \left( \varphi^{-1}(\rho) - \psi_f^{opt}(\rho) \right) {\rm{d}} \rho.
\EE
Fig. \ref{Transfer_GAMP} shows an example of $a_f$ that is illustrated by the shadowed area. The maximal rate of $\psi_f^{opt}$ in \eqref{GAMP-psi-opt} after compression is given by
\BE \label{area-fac}
R_{\text{AC}} = \frac{1}{2\delta} \int_0^{\infty} \psi_f^{opt}(\rho) d\rho.
\EE
Substituting \eqref{AreaPhiG} and \eqref{area-fac} into \eqref{areaf} yields
\BE
a_f =  2\delta \cdot I(y;x) - 2\delta \cdot R_{\text{AC}} \Rightarrow R_{\text{AC}} = I(y;x) - a_{f} / 2\delta. \\ \label{Eqn:Rcs-GAMP}
\EE
The theorem below shows that the gap $a_{f} / 2\delta$ vanishes when $\delta, R_{\text{C}}\to0$ with $R_{\text{C}}/\delta$ fixed.

\begin{theorem} \label{Theom4}
Assume that the matching condition \eqref{GAMP-psi-opt} holds for a symmetrical constellation $\mathcal{S}_C$. Then, $R_{\text{AC}} \to I(y;x)$ when $\delta, R_{\text{C}}\to0$ with $R_{\text{C}}/\delta$ fixed.
\end{theorem}
\begin{IEEEproof}
See Appendix \ref{rate:GAMP}.
\end{IEEEproof}

Theorem \ref{Theom4} shows that $R_{AC}$ of the generalized compressed-coding scheme with a practical $\mathcal{S}_C$ can approach the mutual information between $x$ and $y$ in \eqref{Eqn:SysClip}. To approach $I(y;x)$ in practice, a code should be designed to meet the matching condition \eqref{GAMP-psi-opt} and meanwhile the code rate should be kept as low as possible. However, it is complicated to design a low-rate code to ensure \eqref{GAMP-psi-opt} using curve matching \cite{Yue07,Divsalar11,Ping03}. The same obstacle happens to Theorem \ref{Theom2} as discussed in Remark \ref{remark1}. Indeed, Theorem \ref{Theom2} is a special case of Theorem \ref{Theom4} since $I(y;x) = C_G$ when $f$ in \eqref{Eqn:SysClip} is removed. In the next section, we will treat this issue together.
\section{Analog Spatial Coupling} \label{SecSCCS}
In this section, analog spatial-coupling is introduced to avoid the difficulty in curve matching, which provides a simple method to approach capacity without complicated code optimizations \cite{Yue07,Divsalar11}. Incidentally, we will see in Section \ref{Sec:multiuseer} that SC-CC reveals a new multi-user scheme with short per-user block length and good performance.
\subsection{Spatial Coupling Principle}
\begin{figure}[!htbp]
\centering
  \includegraphics[width=0.45\textwidth]{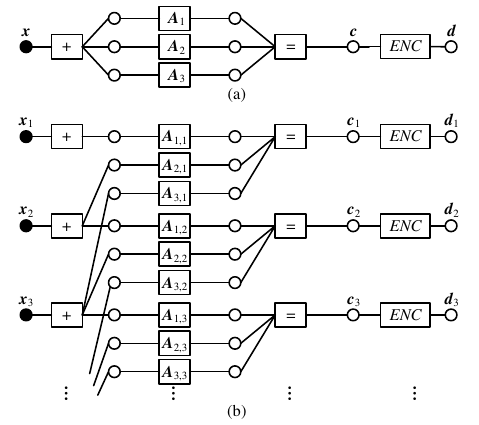}
  \vspace{-0.5cm}
  \caption{(a) A modified version of Fig. \ref{system}(a) with $W = 3$. (b) An SC-CC system based on (a) with $W = 3$.} 
  \label{SC-Structure} \vspace{-0.5cm}
\end{figure}
Fig.~\ref{SC-Structure}(a) shows a slightly modified form of Fig. \ref{system}(a), in which $\ibc$ is repeated for $W$ times ($W=3$ in Fig.~\ref{SC-Structure}(a)). Each replica is multiplied by a compression matrix {$\ibA_i$}. The transmitted signal is
\BE
\ibx=(\ibA_1+\ibA_2+\cdots+\ibA_W)\ibc.
\EE
When the entries of $\ibA$ and {$\{\ibA_i\}$} are i.i.d. Gaussian distributed, Figs.~\ref{system}(a) and \ref{SC-Structure}(a) are equivalent by setting $\ibA= \ibA_1+\ibA_2+\cdots+\ibA_W$ after proper normalization.

Applying the spatial-coupling principle \cite{Kudekar13,Mitchell15,Kudekar15} to $K$ copies of Fig.~\ref{SC-Structure}(a), we obtain an spatially-coupled compressed-coding (SC-CC) scheme in Fig.~\ref{SC-Structure}(b) \cite{Liang16,sLiang17,cLiang17}. The coupling is ``analog'' in that the summation is on the real or complex field. Specifically, for the $j$-th ($1\le j\le K+W-1$) copy, the transmitted signal is given by
\BE \label{SCCSsys}
\ibx_j  = \frac {1} {W} \sum\limits_{k = 1}^K {\ibA_{j+1-k,k} } {\ibc}_{k},
\EE
where $\{\ibA_{j+1-k,k}\}$ are assumed to be independent and the entries of each $\ibA_{j+1-k,k}$ have the same distribution as $\ibA$ in \eqref{AWGNCode}. We assume the termination as $\ibA_{j+1-k,k}=\mathbf{0}$ for $j+1-k<1$ or $j+1-k>W$. When $K$ is large, we approximately have $\mathrm{E}[|x|^2]=1$ if we ignore the boundary effects for $j < W$ and $j > K$. Consider transmitting $\ibx_j$ over an AWGN channel as
\beq \label{SCCS-linear}
\iby_j = \ibx_j + \ibn_j = \frac {1} {W} \sum\limits_{k = 1}^K {\ibA_{j+1-k,k} } {\ibc}_{k}  + \ibn_j,
\eeq
where $j = 1,2,...,K+W-1$ and $\ibn_j$ contains i.i.d. Gaussian noise. The AMP algorithm can be directly applied to the system in \eqref{SCCS-linear}, of which the details can be found in \cite{Barbier17AMP-decoder,Biyik17,Barbier17GAMP}.
\subsection{State Evolution for the SC-CC System} \label{Sec:analysis}
Compared with the SE in \eqref{SE-AMP} for compressed-coding, SE for the SC-CC system is a recursion between two vectors denoted as $\{ v_j, j = 1, ..., K+W-1\}$ and $\{ \rho_k, k = 1, ..., K \}$. Initializing with $\{ v_j^1 = 1,\forall j \}$, the vector SE given by \cite[Definition 4.9]{Barbier17GAMP} is equivalent to the following recursion
\BS \label{EqnSEcoupled}
\begin{eqnarray}
\rho _k^t  \d4&=&\d4 \frac{1}{W}\sum\limits_{w = 1}^W {\phi \left( {v_{k - 1 + w}^t } \right)}, \quad \forall k, \\
v_j^{t + 1} \d4&=&\d4 \frac{1}{W}\sum\limits_{w = 1}^W {\psi \left( {\rho _{j - W + w}^t } \right)}, \quad \forall j,
\end{eqnarray}
\ES
where $\rho _{k}^t  = 0$ for $k<1$ or $k>K$ due to the assumed termination. Functions $\phi(v)$ and $\psi(\rho)$ are the same as those in \eqref{SE-AMP}.
\subsection{Potential Function Analysis}\label{Potential}
Following \cite{Yedla14}, we define the uncoupled potential function as\footnote{Function $U(v)$ is defined based on the underlying uncoupled system in Section \ref{SecUncoupled}. The definition here differs from \cite[Equation (4)]{Yedla14} by an additive constant. Such difference does not affect the minimizer of $U(v)$.}
\BE
U(v) = \int_{\phi(1)}^{\phi(v)} \left[\psi(\rho) - \phi^{-1}(\rho) \right] d\rho, \textrm{  for  } v \in [0,1].
\EE
\begin{lemma}\cite[Theorem 1]{Yedla14}\label{Lem2}
For the coupled recursion in \eqref{EqnSEcoupled}, the variance of the fixed point is upper bounded by the minimizer of $U(v)$ when $K,W\to\infty$.
\end{lemma}
\par
Here, we assume that $U(v)$ has a unique minimum. From Lemma \ref{Lem2}, a sufficient condition for error-free decoding is that the minimizer of $U(v)$ tends to zero.

\begin{figure}[!htbp]
\centering
\subfigure[]{
\begin{minipage}[t]{0.3\textwidth}
\centering \label{MultiCross}
\includegraphics[width=1\textwidth]{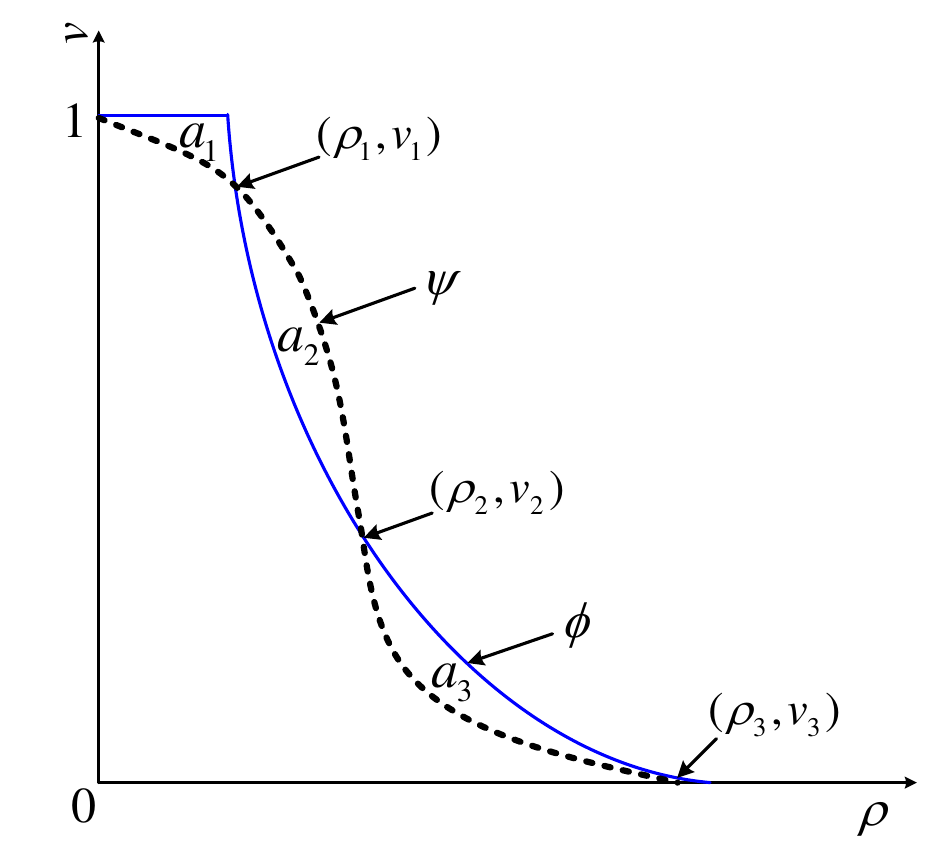}
\end{minipage} }
\subfigure[]{
\begin{minipage}[t]{0.3\textwidth}
\centering \label{SingleCross}
\includegraphics[width=1\textwidth]{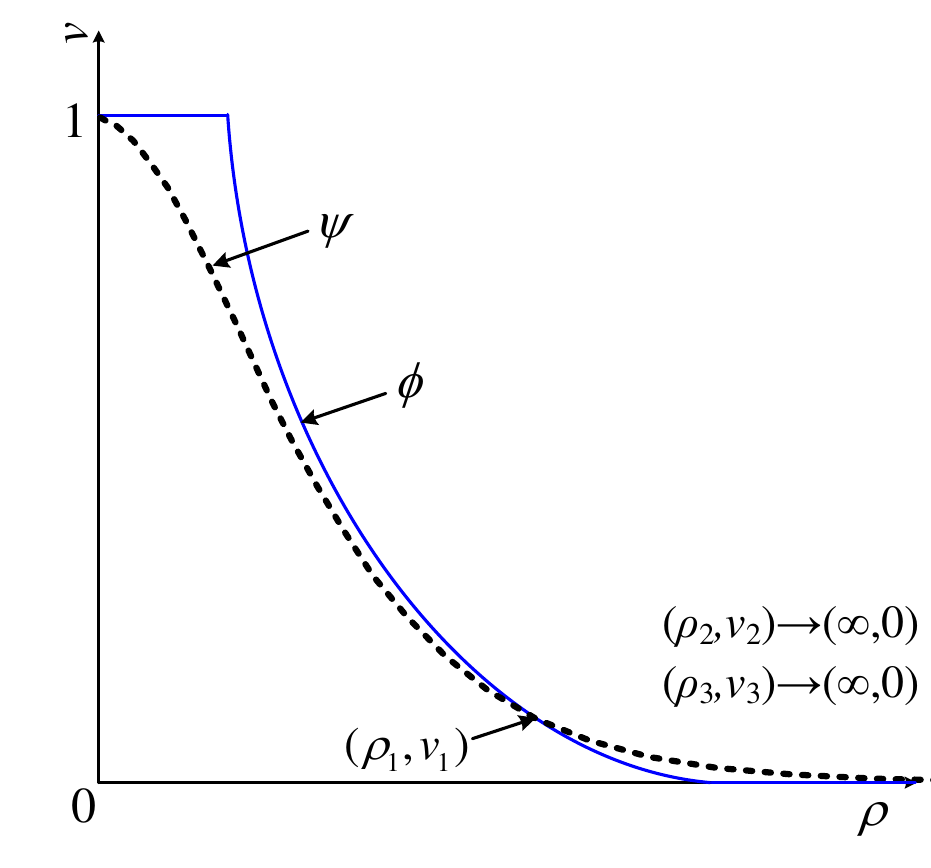}
\end{minipage}}
\caption{Examples of $\phi$ and $\psi$ in typical scenarios.}\label{Transfer2} \vspace{-0.3cm}
\end{figure}

\revise{We consider typical scenarios of $\phi$ and $\psi$ in Fig. \ref{Transfer2} to illustrate Lemma \ref{Lem2}. In Fig. \ref{MultiCross}, $\phi$ and $\psi$ have three intersections marked as {$(\rho_i,v_i)$}, $i=1, 2, 3$. In particular, we assume that $v_3 \approx 0$, which implies error free decoding approximately. Fig. \ref{SingleCross} can be seen as an special case of Fig. \ref{MultiCross}, where ($\rho_2, v_2$) $\to$ ($\infty, 0$) and ($\rho_3, v_3$) $\to$ ($\infty, 0$).} The areas $a_1$, $a_2$ and $a_3$ in Fig. \ref{MultiCross} are, respectively, calculated as
\BE \label{a1-a3}
a_1 \equiv \int_{v_1}^1 {\left[ {\phi \left( v \right){\rm{ - }}\psi ^{ - 1} \left( v \right)} \right]{\rm{d}}} v, \quad a_2 \equiv \int_{v_{\rm{2}} }^{v_{\rm{1}} } {\left[ { \psi ^{ - 1} \left( v \right) - \phi \left( v \right) } \right]{\rm{d}}} v, \quad a_3 \equiv \int_{v_3}^{v_2} {\left[ {\phi \left( v \right){\rm{ - }}\psi ^{ - 1} \left( v \right)} \right]{\rm{d}}} v.
\EE
It can be verified that the minimizer of $U(v)$ is either $v_1$ or $v_3$. The critical point is $U(v_1)=U(v_3)$, or equivalently, $a_2=a_3$. According to Lemma \ref{lemma2} and \eqref{a1-a3}, we have from Fig. \ref{MultiCross}
\BE
R_{C} = \frac{1}{2}\int_0^\infty  {\psi \left( \rho  \right){\rm{d}}} \rho = \frac{1}{2} \left\{ \int_0^1 {\phi \left( v \right){\rm{d}}v} - a_1 + a_2 - a_3 \right\}.
\EE
When the critical condition ($a_2=a_3$) is reached, the achievable rate $R_{AC}$ is given by
\BE \label{ascrate}
R_{AC} = \frac{R_{C}} {\delta} = \frac{1}{2\delta} \left\{ \int_0^1 {\phi \left( v \right){\rm{d}}v} - a_1 \right\} = C_{\text{G}} - a_{1} / 2\delta,
\EE
where the termination effects are ignored. For the special case in Fig. \ref{SingleCross}, according to \eqref{a1-a3}, the critical point is given by $a_2 \to 0$ and $a_3 \to 0$.


\revise{\subsection{Achievable Rate of SC-CC}
\begin{figure}[!htbp]
\centering
\subfigure[SNR = 0.5 dB and $C_G = 0.54$]{
\begin{minipage}[t]{0.48\textwidth}
\centering \label{SE0dB}
\includegraphics[width=1\textwidth]{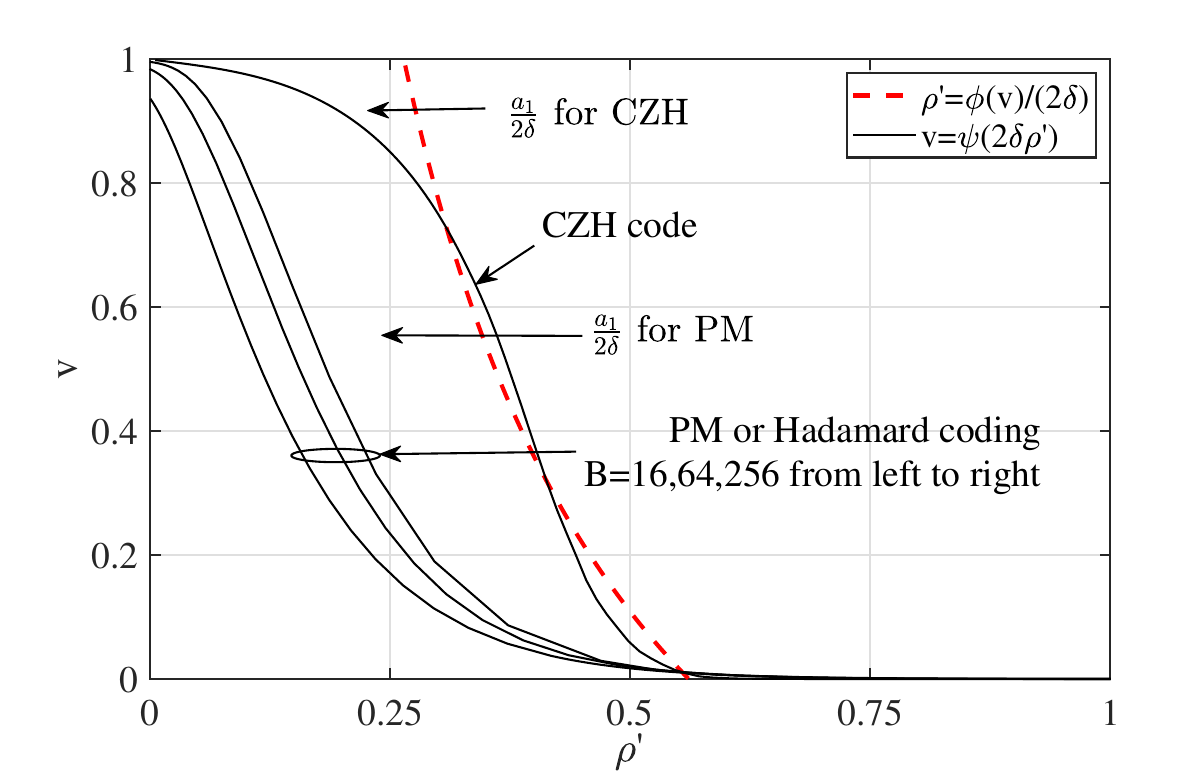}
\end{minipage} }
\subfigure[SNR = 15 dB and $C_G = 2.5$]{
\begin{minipage}[t]{0.48\textwidth}
\centering \label{SE15dB}
\includegraphics[width=1\textwidth]{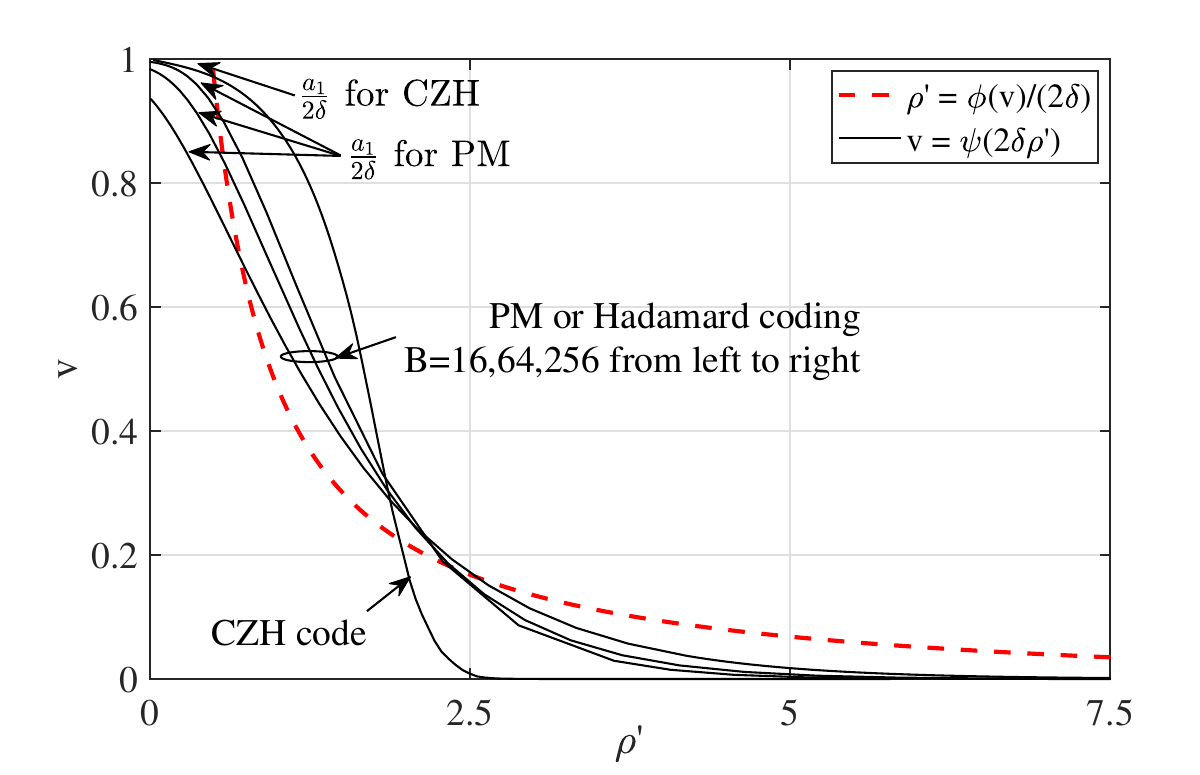}
\end{minipage}}
\caption{Transfer functions $\phi$ and $\psi$ for different underlying codes. The CZH code (non-systematic and un-punctured) has two component codes  with BPSK modulation and the Hadamard code length is 64 \cite{Leung06}. (a) $\delta$ = 1.1977, 0.3628, 0.1065 for PM with $B$ = 16, 64, 256, respectively. $\delta$ = 0.0916 for CZH.  (b) $\delta$ = 0.1193, 0.0429, 0.0138 for PM with $B$ = 16, 64, 256, respectively. $\delta$ = 0.0205 for CZH. } \label{SE:SR}
\vspace{-0.5cm}
\end{figure}
The critical condition $a_2=a_3$ for \eqref{ascrate} can be equivalently stated as
\BE \label{solveDelta}
\int_{0}^{v_1} \phi(v) dv = \int_{0}^{v_1} \psi ^{-1}(v) dv, 
\EE
where $v_1 \in (0,1)$ is the root of $\phi(v)=\psi ^{-1}(v)$. Recall that $\phi(v) = \frac{\delta}{v+\sigma^2}$, implying that $v_1$ is a function of $\delta$. Hence, the critical condition in \eqref{solveDelta} determines a unique value for $\delta$, which in turn determines the achievable rate after compression, i.e., $R_{AC}=\frac{R_C}{\delta} = \frac{1}{2\delta} \int_{0}^{\infty}\psi(\rho)d\rho$.}

\revise{Fig. \ref{SE:SR} illustrates the curves $\rho'=\phi(v)/2\delta$ and $v=\psi(2\delta\rho')$ for PM with different $B$ and a concatenated zigzag Hadamard (CZH) code \cite{Leung06}. The curves for PM are identical to those for Hadamard codes, as analyzed in Section \ref{Equivalence}. More details on the CZH code will be given in Section \ref{SectSim1}. Fig. \ref{SE:SR} involves scaling $\rho' = \frac{\phi(v)}{2\delta} = \frac{0.5}{v+\sigma^2}$  to make $\rho'$ be independent of $\delta$. The values of $\delta$ for all $\psi(2\delta\rho')$ curves are numerically computed such that \eqref{solveDelta} is satisfied. According to \eqref{Eqlemma1}, \eqref{Eqn:lemma2} and \eqref{ascrate}, we have
\BE \label{areaSCscaled}
\int_0^1 \phi(v)/(2\delta) dv = C_G \quad \mathrm{and} \quad \int_{0}^{\infty}\psi(2\delta\rho')d\rho' = R_{AC} = C_G - a_1/(2\delta),
\EE
where the areas indicated by $\frac{a_{1}}{2\delta}$ in Fig. \ref{SE:SR} give the gaps to the capacity.}
	
For the relatively low SNR in Fig. \ref{SE0dB}, the achievable rate of SC-SRC is quite poor. This is because the slow roll-off rates of the PM curves, which lead to large values of $\delta$ from (56) and hence small values of $R_{AC}=R_{C}/\delta$. SC-SRC performs better at a high SNR in Fig. \ref{SE15dB}. The reason is that the flatter curves of $\rho' = \frac{0.5}{v+\sigma^2}$ lead to smaller values of $\delta$ and hence larger values of $R_{AC}$. We can see from Fig. \ref{SE:SR} that the CZH code performs better as its curves have high roll-off rates in both low and high SNRs, resulting in quite small values of $\delta$. Incidentally, the roll-off rate of PM increases when $B$ increases. This is consistent of the claim that SC-SRC is asymptotically capacity approaching when $B \to \infty$. However, decoding cost can be a concern when $B$ is large.
	
\revise{The above analyses provide useful insights into the rationales of SC-SRC and SC-CC. As a final note, Theorems \ref{Theom2} and \ref{Theom4} are under the assumption that $\psi(\rho)$ in the vicinity of $\rho=0$ is given by symbol-by-symbol estimation. A classic non-concatenated code, such as PM or a Hadamard code, generally does not meet this assumption. A concatenated code fits this assumption better. The latter usually has performance close to symbol-by-symbol estimation before a certain SNR threshold and a sharp water-fall behavior afterward \cite{Mitchell15}.}
\section{Simulation Results}
In this section, we provide simulation results to show the advantages offered by SC-CC in approaching Gaussian capacity, universal coding and short block length coding in multi-user systems.
\subsection{Compressed-Coding with a Low Rate Underlying Code}\label{SectSim1}
\subsubsection{Low-Rate CZH Coding}\label{SectSim1}
A CZH code (non-systematic and un-punctured) with BPSK modulation is used for $\ibc$ with two component codes \cite{Leung06}. The Hadamard code length is 64, the number of information bits (per copy) is $24576$ and $R_C= 3/64$. For complexity considerations, the compression matrix is generated using randomly selected rows from an $N \times N$ Hadamard matrix. We observed that the difference between Hadamard and i.i.d. Gaussian sensing matrices is typically very small for large $N$. 
Fast Hadamard transform~(FHT) is used with complexity $\log_2(N)$ per bit. Soft output FHT is used for decoding the CZH code \cite{Leung06}. Iteration proceeds until convergence.

Theoretically, AMP requires asymptotically large sensing matrices, which incurs high cost even with FHT. Simulations show that the performance of the above scheme remains almost unchanged for $N > 65536$. Therefore, to reduce cost, each copy~(coded length $= 524288$) is partitioned into $8$ parts, each of length $65536$, that are individually compressed.
\begin{figure}[!t]
\centering
\subfigure[$R_{\text{AC}} = 0.5$ and $R_{\text{SC}} = 0.4808$]{
\begin{minipage}[t]{0.48\textwidth}
\centering \label{SimulationV1-a}
\includegraphics[width=1\textwidth]{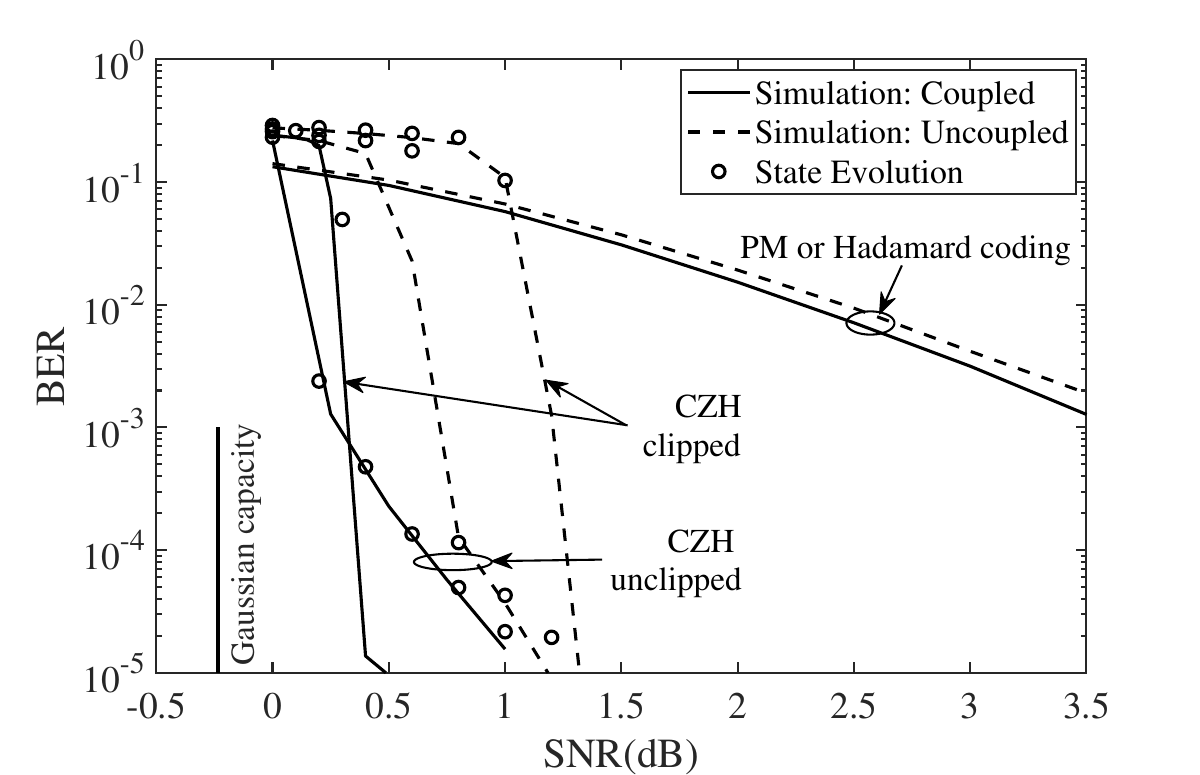}
\end{minipage} }
\subfigure[$R_{\text{AC}} = 1.0$ and $R_{\text{SC}} = 0.9615$]{
\begin{minipage}[t]{0.48\textwidth}
\centering \label{SimulationV1-b}
\includegraphics[width=1\textwidth]{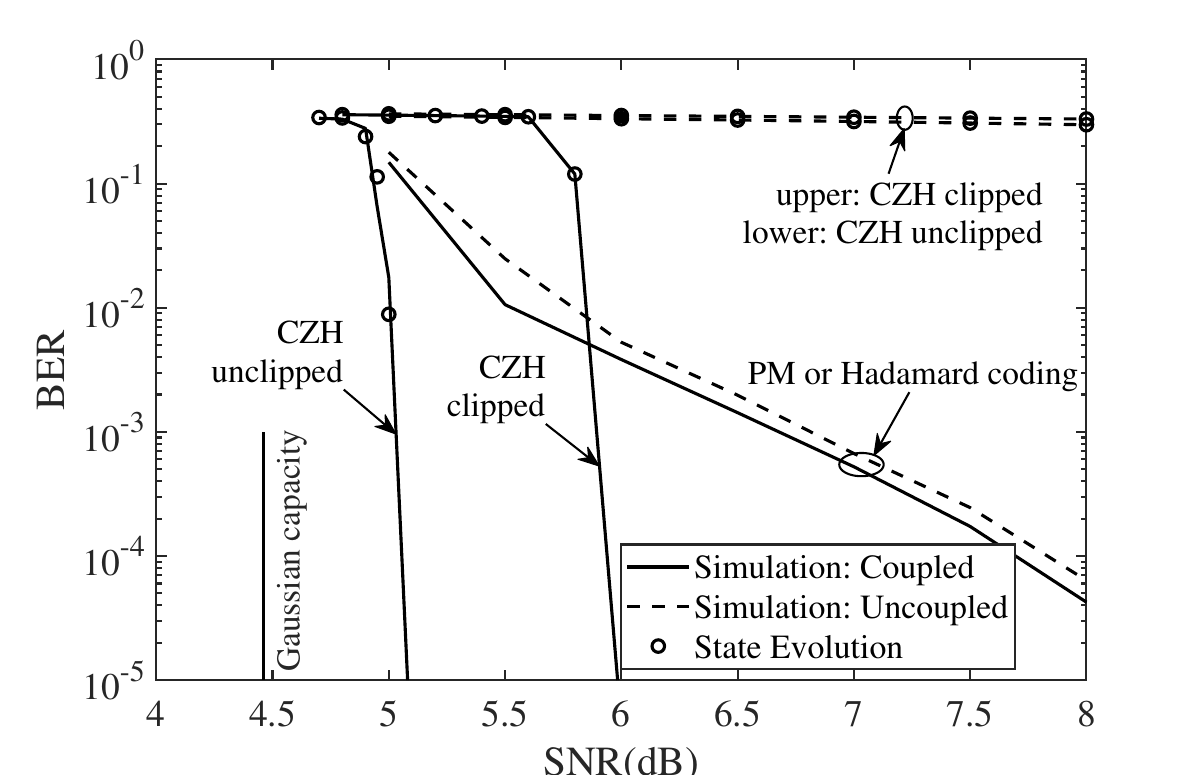}
\end{minipage}}
\caption{Simulation results for compressed-coding. PM and Hadamard coding refer to sparse regression coding and compressed-coding, respectively. $L = 4096$ and $B = 128$ for PM. $K=50$ and $W=3$ for spatial-coupling. CR = 1 dB for clipping.} \label{SimulationV1}
\end{figure}
\subsubsection{Spatial-Coupling and Clipping}
Recall that the theoretical analysis for spatial-coupling in Section \ref{Potential} requires $K, W\to\infty$. For practical $K$ and $W$, termination incurs rate loss in spatial-coupling and the actual rate realized by SC-CC is given by $R_{\text{SC}} \equiv R_{\text{AC}}\cdot K / (K+W-1)$ \cite{Kudekar13}. \revise{Fig. \ref{SimulationV1} shows the bit error rate (BER) performance of compressed-coding in various settings. For SC-CC with clipping, the received signal is given by
\BE\label{EqnClipAWGN}
\iby_j =\alpha \cdot \mathrm{clip}(\ibx_j)  + \ibn_j, \quad j = 1,2,...,K+W-1,
\EE
where $\mathrm{clip}(\cdot)$ and $\{ \ibx_j, \forall j \}$ are given by \eqref{Eqnclip} and \eqref{SCCSsys}, respectively. The coefficient $\alpha \triangleq \sqrt{1/ \mathrm{E} [ |\mathrm{clip}(x)|^2]}$ normalizes the transmit power to unit. Define the clipping ratio (CR) as $\mathrm{CR}\triangleq 10 \log_{10}(Z^2/\mathrm{E} [x^2])$.}

\revise{From Fig. \ref{SimulationV1}, we have the following observations:
\begin{itemize}
  \item The SE predictions are reasonably accurate for AMP or GAMP based decoding. Note that SE assumes i.i.d. Gaussian matrices while simulations use the Hadamard matrices due to complexity concerns.
  \item Spatial-coupling offers significant performance gains for compressed-coding. Such gains are marginal for sparse-regression code at low-to-medium SNRs. For the latter, the gain is more prominent at high SNRs \cite{Barbier17AMP-decoder,Rush20,Hsieh20}.
  \item SC-SRC and SC-CC based on Hadamard codes have exactly the same BER performance. Similar results are reported in \cite{cLiang17}. SC-CC based on a CZH code offers excellent performance. This is consistent with the analysis related to Fig. \ref{SE:SR}.
  \item Clipping causes performance loss in the uncoupled cases. However, very interestingly, clipping may improve performance in the coupled case. This phenomenon was first reported in \cite{sLiang17}, where an intuitive explanation is given based on the area property. The improvement is significant at $R_{SC}=0.48$: SC-CC with clipping approaches about 0.7 dB away from the Gaussian capacity.
\end{itemize}}
\subsection{Universal Coding Scheme} \label{SubSectModified}


\begin{figure}[!htbp]
\centering
  \includegraphics[width=0.45\textwidth]{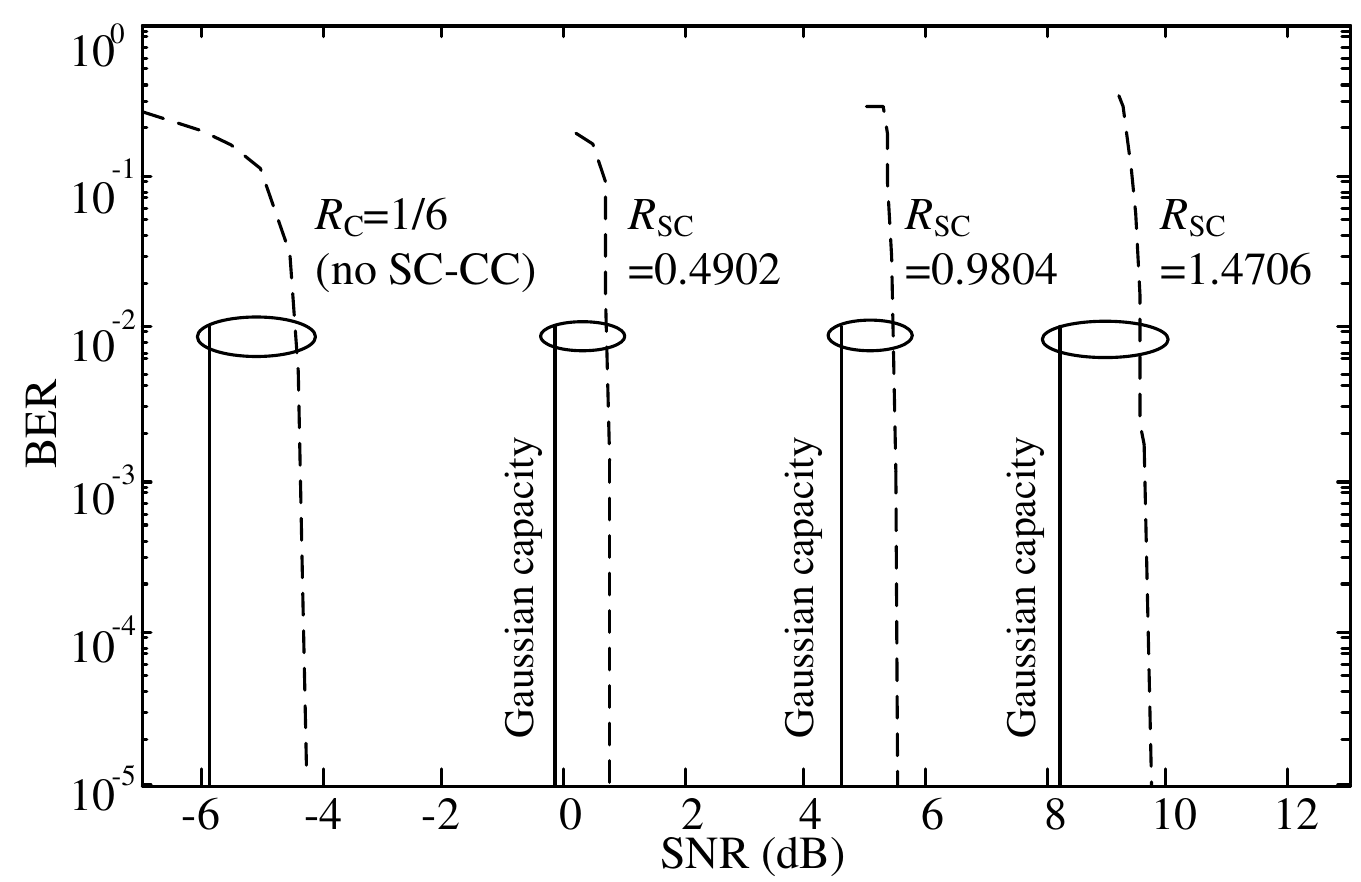} \vspace{-0.5cm}
  \caption{Performance of the modified SC-CC system.}\vspace{-0.5cm}
  \label{SimulationV2}
\end{figure}

We consider a modified version of Fig. \ref{SC-Structure}(b), in which each input $\ibd_k$ is, after independent interleaving, encoded for $W$ times. Graph illustration can be found in \cite[Fig. 5]{Liang16}. We observed that the extra interleaving leads to performance improvement. However, so far, we are unable to provide analytical explanations. Fig. \ref{SimulationV2} shows the BER performance of the modified SC-CC system using a rate-$1/2$ non-systematic and un-punctured ZH code~\cite{Leung06} with Hadamard code length $= 4$ and number of information bits per copy $=16384$. The rate of the CZH code $= 1/6$. Each copy is partitioned into $3$ parts that are individually compressed. The sensing matrices are based on a size $N_{H} = 32768$ Hadamard matrix. For spatial-coupling, $W = 3$ and $K = 100$.

We call this SC-CC scheme as a master code. We can puncture this master code to obtain different rates. Fig. \ref{SimulationV2} shows the simulation results for $R_{\text{SC}} = 0.4902$, $0.9804$ and $1.4706$, where random puncturing is used. The non-puncturing rate is $1/6$. From Fig. \ref{SimulationV2}, we can see that this master code is universal since it can be randomly punctured without affecting the relative performance measured by the gaps toward capacity. At $R_{\text{SC}}=0.4902$, nearly error-free performance is achieved at only $0.7$ dB away from the theoretical limit for Gaussian signaling. Such universal codes have been widely discussed for various applications, such as type-II ARQ \cite{Huang07,Zhang09} and distributed caching systems \cite{Borst10}.


\subsection{A Multiuser SC-CC System with Short Block Length per User}\label{Sec:multiuseer}
We now consider the application of SC-CC in a multi-user system of $K$ users. The overall system structure is the same as that in Fig. \ref{SC-Structure}(b) except that $\{ \ibd_k \}$ are generated separately by $K$ users. The transmitted signals from $K$ users are encoded and transmitted in a decentralized way without information sharing except for proper transmission time scheduling. Specifically, we divide the time span into slots, each of which corresponds to a copy. In slot $k$, user $k$ generates the signals based on its data $\ibd_k$ and transmits them over $W$ slots starting from $k$. The signals from different users are separated by user-specific interleaving, following the interleave division multiple access~(IDMA) principle~\cite{Ping06}.  At the receiver, the signals from $K$ users are combined, which has the same effect as a linear summation in Fig. \ref{SC-Structure}(b) \cite{Truhachev13a}.

In general, to achieve improved performance, a spatially-coupled LDPC code requires much longer block length than the underlying LDPC code before coupling \cite{Kudekar15,Mitchell15}. This is because the former involves multiple copies of the latter. In a multiuser SC-CC system, the increased overall block length is shared by multiple users. This achieves the benefit of spatial-coupling without increasing the codeword length of each user.

Incidentally, the block length problem is usually due to the latency constraint. It cannot be solved by, e.g., increasing processing speed. It is a source problem, and in many real-time applications, the source can only generate a limited number of information bits within a fixed duration. The scheme below is to fill this fixed duration with the signals from multiple users, which effectively increases block length.

\begin{figure}[!htbp]
\centering
  \includegraphics[width=0.45\textwidth]{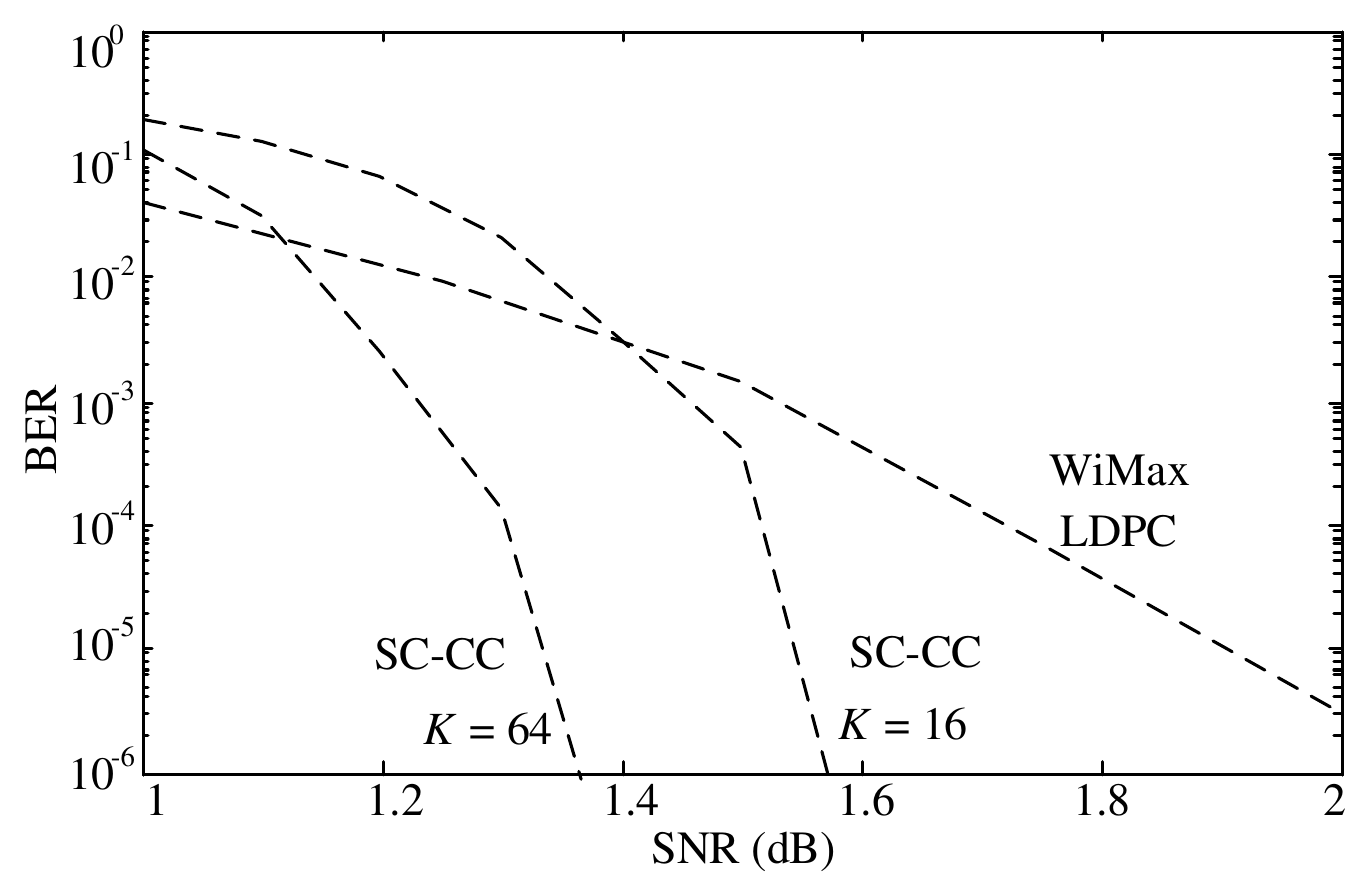}\vspace{-0.5cm}
  \caption{Multiuser performance of the modified SC-CC scheme and the WiMax LDPC code~\cite{Tal15}. All schemes have a coding rate $0.5$.}
  \label{SimulationV3}\vspace{-0.5cm}
\end{figure}

Fig.~\ref{SimulationV3} shows the performance of the SC-CC-IDMA scheme for both $K = 16$ and $64$ using a rate-$1/4$ non-systematic and un-punctured ZH code with Hadamard code length = 16. The rate of the CZH code = 1/16. Number of information bits (per copy) = 1024. Each copy is partitioned into 4 parts that are individually compressed. To achieve a rate exactly $R_{\text{AC}} = 0.5$, for $K = 16$, each $\ibA_{j,w}$ is formed by randomly selecting $1725$ rows in an Hadamard matrix with $N_H = 4096$ so $\delta = 0.4211$; for $K = 64$, each $\ibA_{j,w}$ is formed by randomly selecting $1956$ rows in an Hadamard matrix with $N_H = 4096$ so $\delta = 0.4775$. For spatial-coupling, $W = 4$. The SC-CC performance improves with $K$. This is a common property of spatial-coupling~\cite{Mitchell15}, since the overhead due to termination reduces when $K$ increases.

The performance of the rate-$1/2$ LDPC code for the WiMax standard~\cite{Tal15} with $1152$ information bits is compared in Fig.~\ref{SimulationV3}, for which users are separated by time division multiple access (TDMA). We can see that SC-CC outperforms the conventional LDPC coded TDMA scheme noticeably. Intuitively, the coding rate of each user in SC-CC is $1/8$~(after compression) since it occupies $W = 4$ slots.
This is much lower than the rate ($=1/2$) of the LDPC code. Therefore, SC-CC might provide better coding gain if cross user interference can be ignored. However, interference does exist in SC-CC. It appears that SC-CC provides an efficient way for multiuser interference cancelation. This phenomenon was first noted in~\cite{Yedla11} for LDPC codes with SC but without compression.

CDMA with conventional successive interference cancelation~(SIC) cannot help in this case. If a short code is used by each user in CDMA, decoding loss will accumulate in SIC, resulting in large overall loss. Superposition coding~\cite{Ma04} also suffers from the same problem. Clearly, SC-CC offers an attractive solution to the latency problem by sharing a long block length by multiple users in multi-user systems.

\section{Conclusions}\vspace{-0.3cm}

In this paper, we proposed a compressed-coding scheme that combines compressive sensing with FEC coding, where AMP is used for decoding. We derived the performance limit of compressed-coding and showed that compressed-coding can asymptotically approach Gaussian capacity. This capacity approaching property can be maintained in systems with non-linear effects such as clipping and quantization. We also studied an SC-CC scheme to circumvent the difficulty in optimizing low-rate codes for approaching capacity in compressed-coding. We showed that SC-CC can maintain universally good performance under random puncturing. By sharing the overall code length among multiple users, SC-CC also relieves the requirement on per-user block length to achieve good performance in multi-user environments. The above claims are supported by extensive theoretical and numerical results.
\begin{appendices}\vspace{-0.3cm}
\section{Proof of Theorem \ref{delta/a2zero1}}\vspace{-0.3cm} \label{proofdelta/a2zero1}
It can be verified that $\text{mmse}(\mathcal{B},\rho) = \phi^{-1}(\rho)$ has exactly one solution for $\rho \in [0,\infty)$, which is denoted as $\rho_{\mathcal{B}}$. Recalling $\rho = \phi(v)=\frac{\delta}{v+\sigma^2}$, we have $\rho \to 0$ as $\delta \to 0$. Apply the first-order Taylor expansion to $\text{mmse}(\mathcal{B},\rho)$ in \eqref{Eqn:MMSE_b} at $\rho = 0$ as \cite{Guo2011}
\BE \label{Eqn:MMSE_app}
v = \text{mmse} \left( \mathcal{B}, \rho  \right) = 1 - \rho  + o\left( \rho^2  \right).
\EE
Substituting \eqref{Eqn:MMSE_app} into $\phi(v)$ yields
\BS
\begin{eqnarray}
\rho_{\mathcal{B}} = \frac {\delta} {1 - \rho_{\mathcal{B}} + \sigma^2 } \Longleftrightarrow \d4\d4&& \rho_{\mathcal{B}}^2 - (1+\sigma^2) \rho_{\mathcal{B}} = - \delta \\
\Longleftrightarrow \d4\d4&& \left( \rho_{\mathcal{B}} - \frac {1+\sigma^2} {2} \right)^2 = \left( \frac {1+\sigma^2} {2} \right)^2  -  \delta \\
\overset{(a)}{\Longleftrightarrow} \d4\d4&& \rho_{\mathcal{B}} = \frac {1+\sigma^2} {2} - \sqrt{\left( \frac {1+\sigma^2} {2} \right)^2  -  \delta}, \label{rhoB}
\end{eqnarray}
\ES
where $(a)$ is due to that we consider the solution at the vicinity of $\rho = 0$ and the other solution (a larger value) is abandoned.

According to \eqref{MatchingBPSK}, we have
\BE \label{psi-BPSK}
\psi_{\mathcal{B}}^{opt}(\rho) = \left\{
       \begin{array}{ll}
       \text{mmse}(\mathcal{B},\rho), & \hbox{$0 \leq x \leq \rho_{\mathcal{B}}$,} \\
       \phi^{-1}(\rho), & \hbox{$\rho_{\mathcal{B}} < x$.}
       \end{array}
     \right.
\EE
For $a_{\mathcal{B}}$ in \eqref{as_BPSK}, it can be shown from \eqref{psi-BPSK} that
\beq
a_{\mathcal{B}} = \int_0^{\rho_{\mathcal{B}}} { \left( \phi^{-1}(\rho) - \text{mmse}  (\mathcal{B}, \rho)  \right){\rm{d}}} \rho \le \int_0^{\rho_{\mathcal{B}}} {\left(1 - \text{mmse} ( \mathcal{B}, \rho)  \right){\rm{d}}} \rho. \label{Eqn:Delta_bound}
\eeq
Combining \eqref{Eqn:Delta_bound} and \eqref{Eqn:MMSE_app} yields
\BE \label{areaBPSK}
a_{\mathcal{B}}  \le \int_0^{\rho_B} \left( \rho + o (\rho^2)  \right) \rm{d} \rho =0.5\rho_{\mathcal{B}}^2+o(\rho_{\mathcal{B}}^3).
\EE
In \eqref{rhoB}, we treat $\rho_{\mathcal{B}}$ as a function of $\delta$ and apply Taylor expansion at $\delta = 0$ as
\BE \label{Eqn:fixed_point}
\rho_{\mathcal{B}}  = \frac{\delta }{{1 + \sigma ^2 }} + o\left( \delta  \right).
\EE
Substituting \eqref{Eqn:fixed_point} into \eqref{areaBPSK}, we have
\BE \label{Eqn:Delta_asym}
\frac{a_{\mathcal{B}}}{2\delta } \le 0.25\frac{\delta }{{\left( {1 + \sigma ^2 } \right)^2 }} + o\left( \delta  \right).
\EE
As $\delta \rightarrow 0$, we can see from \eqref{Eqn:Delta_asym} that $a_{\mathcal{B}}/ 2\delta \rightarrow 0$, and therefore complete the proof.
\vspace{-0.3cm}\section{Proof of Theorem \ref{Theo1}}\vspace{-0.3cm} \label{SectProof}
Start with evaluating $I(S_1;S_3)$ as
\BS
\begin{eqnarray}
I(S_1;S_2,S_3) = I(S_1;S_3,S_2) \overset{(a)}{\Longleftrightarrow} \d4\d4 && I(S_1;S_3)+I(S_1;S_2|S_3)=I(S_1;S_2)+I(S_1;S_3|S_2) \label{rearrange} \\
\overset{(b)}{\Longleftrightarrow} \d4\d4 && I(S_1;S_3) = I(S_1;S_2) - I(S_1;S_2|S_3). \label{I_phat_z}
\end{eqnarray}
\ES
In \eqref{rearrange}, we simply rearrange the variables.
In ($a$), we apply the chain rule of the conditional mutual information.
In ($b$), we apply the property of Markov chain, i.e., $I(S_1;S_3|S_2)=0$. From \eqref{I_phat_z}, we evaluate $I(S_1;S_2)$ and $I(S_1;S_2|S_3)$  separately as follows.

Based on interpretation in \cite{Guo2011}, $S_1$ and $S_3$ can be viewed as two independent observations for $S_2$. In particular, $S_1$ is an AWGN observation of $S_2$ according to \eqref{joint_z_phat} and \eqref{Cov_z_phat}. From \cite{Guo2011}, we have $I(S_1;S_2)=0.5 \text{log}(1+\rho)$ with $\rho$ being the effective SNR of the AWGN channel. Following the mutual information and MMSE identity for AWGN channel with side information, we have \cite{Guo2011}
\BE
I(S_1;S_2|S_3=s_3)= \frac{1}{2}\int_0^{\rho} \mathrm{mmse}(\rho, S_2|S_3=s_3)d\rho,
\EE
where $s_3$ is a realization of $S_3$ and $\mathrm{mmse}(\rho,S_2|S_3=s_3)$ represents the minimum MSE of estimating $S_2$ from the AWGN observation $S_1$ with side information $S_3=s_3$. Taking expectation of $I(S_1;S_2|S_3=s_3)$ with respect to (w.r.t.) $S_3$ yields
\BE
I(S_1;S_2|S_3) = \mathrm{E} [I(S_1;S_2|S_3=s_3)] = \frac{1}{2} \int_0^{\rho} \mathrm{E} [\mathrm{mmse}(\rho, S_2|S_3=s_3)] d\rho = \frac{1}{2} \int_0^{\rho} \mathrm{mmse}(\rho, S_2|S_3) d\rho. \label{mmse_int}
\EE
For \eqref{I_phat_z}, we substitute $I(S_1;S_2)$ and $I(S_1;S_2|S_3)$ and yield
\BE
I(S_3;S_1) = \frac{1}{2} \text{log}(1+\rho) - \frac{1}{2} \int_0^{\rho} \mathrm{mmse}(\rho, S_2|S_3) d\rho. \label{mutual_rho}
\EE
Note that $\mathrm{mmse}(\rho,S_2|S_3)$ in \eqref{mutual_rho} is a function of the effective SNR $\rho$. Taking partial derivative of $I(S_3;S_1)$ w.r.t. $\rho$ in \eqref{mutual_rho} yields
\beq
\frac {\partial} {\partial \rho} I(S_3;S_1) = \frac {\partial} {\partial \rho}  \lS \frac{1}{2}\text{log}(1+\rho) - \frac{1}{2}\int_0^{\rho} \mathrm{mmse}(\rho, S_2|S_3) d\rho \rS = \frac {1} {2(1+\rho)} - \frac{1}{2}\mathrm{mmse}(\rho, S_2|S_3). \label{derivative}
\eeq

Next, the effective SNR $\rho$ should be identified for the effective AWGN channel. According to \eqref{joint_z_phat} and \eqref{Cov_z_phat}, $p(S_1,S_2)$ is given by
\BE
p(S_2,S_1) \propto \mathrm{exp} \lp - \frac {1} {2} [S_2,S_1] \mathbf{\Sigma}^{-1} [S_2,S_1]^T \rp \propto \mathrm{exp} \lS - \frac {1} {2} \lp \frac {S_1^2} {v(1-v)} - \frac {2} {v} S_2S_1 + \frac {S_2^2} {v} \rp \rS. \label{Py_qua}
\EE
According to Bayes' rule, $p(S_1|S_2)$ is given by
\BS
\begin{eqnarray}
p(S_1|S_2) = \frac{p(S_2,S_1)}{p(S_2)} \d4\d4 && \propto \frac {\mathrm{exp} \lS - \frac {1} {2} \lp \frac {1} {v(1-v)} S_1^2 - \frac {2} {v} S_2S_1 + \frac {1} {v} S_2^2 \rp \rS } {\mathrm{exp} \lS - \frac {S_2^2} {2} \rS} \\
&& \propto \mathrm{exp} \lS - \frac {1} {2} \lp \frac {S_1^2} {v(1-v)} - \frac {2} {v} S_2S_1 + \frac {1-v} {v} S_2^2 \rp \rS \\
&& \propto \mathrm{exp} \lS - \frac {S_1^2 - 2(1-v)S_2S_1 + (1-v)^2 S_2^2 } {2v(1-v)} \rS \\
&& \propto \mathrm{exp} \lS - \frac { \lp S_1 - (1-v) S_2 \rp^2 } {2v(1-v)} \rS.
\end{eqnarray}
\ES
The conditional distribution of $p(S_1|S_2)$ is given by
\BE
p(S_1|S_2) = \mathcal{N} \lp S_1;(1-v)S_2,v(1-v) \rp. \label{p_phat_over_z}
\EE
Eq. \eqref{p_phat_over_z} is equivalent to the following AWGN channel
\BE
S_1 = (1-v) \cdot S_2 + \mathcal{N} \lp 0,v(1-v) \rp,\label{Conditionalzbar}
\EE
where the noise is independent of $S_2$. The effective SNR $\rho$ is given by
\BE
\rho = \frac {(1-v)^2} {v(1-v)} \Leftrightarrow v = \frac {1} {1+\rho}.
\EE
Substituting $v = \frac {1} {1+\rho}$ into \eqref{derivative} yields
\BS
\begin{eqnarray}
-\frac {\partial} {\partial v} I(S_3;S_1) \d4\d4&& = \lS - \frac {\partial} {\partial \rho} I(S_3;S_1) \rS_{\rho = \frac {1-v} {v}} \cdot \frac {d} {dv} \rho \\
\d4\d4&& =\lp \frac {1} {2(1+\rho)} - \frac{\mathrm{mmse}(\rho, S_2|S_3)}{2} \rp_{\rho = \frac {1-v} {v}} \cdot \lp \frac {1} {v^2} \rp \\
\d4\d4&& =\frac{1}{2} [v- \mathrm{mmse}(S_2|S_1,S_3)] \cdot \frac {1} {v^2} \label{GAMP_Fisher} \\
\d4\d4&& = \frac {1} {2v} \lp 1- \frac  {\mathrm{mmse}(S_2|S_1,S_3)} {v} \rp.
\end{eqnarray}
\ES
Therefore,
\BE
- \frac {\partial} {\partial v} I(S_3;S_1) = \frac {1} {2v} \lp 1- \frac  {\mathrm{mmse}(S_2|S_1,S_3)} {v} \rp,
\EE
which completes the proof of Theorem \ref{Theo1}.

\section{} \label{rate:GAMP}
\subsection{An Upper Bound of $\varphi(v)$}
For $\by = f(\ibx)+\ibn$, we assume that $f$ does not change the average power of $\ibx$, i.e., $\mathrm{E}[|f(x)|^2]=\mathrm{E}[|x|^2]$. The following proposition gives an upper bound of $\varphi$.
\begin{proposition}\label{propsition1}
For $v \in [0,1]$, $\rho = \varphi(v)$ in \eqref{GAMPphi} is upper bounded by $\delta \cdot \left( \frac{1+\sigma^2}{4\sigma^2} \right)$.
\end{proposition}
\begin{IEEEproof}
Define $\ibz = f(\ibx)$ and $\textrm{mmse}(z|y)$ as the minimum MSE of estimating $\ibz$ from $\iby=\ibz+\ibn$. Note that $\mathrm{E}[|f(x)|^2]=1$. According to \cite{Guo2011}, $\textrm{mmse}(z|y) \geq \frac{1}{1+1/\sigma^2}$. According to the data processing inequality, it is clear that $\textrm{mmse}(x|y) \geq \textrm{mmse}(z|y)$.

Rewrite $\varphi(v)$ in \eqref{GAMPphi} as
\BE \label{GAMPphi2}
\rho = \varphi(v) = \delta \left( v+ \frac {1} {\frac{1}{\textrm{mmse}(x|\hat{p},y)} - \frac{1}{v} } \right)^{-1}.
\EE
Combining $\textrm{mmse}(x|y) \geq \frac{1}{1+1/\sigma^2}$ with \eqref{GAMPphi2} yields
\BS
\begin{eqnarray}
\rho \leq \delta \left( v+ \frac {1} {1+1/\sigma^2 - \frac{1}{v} } \right)^{-1} \d4& = &\d4 \delta \left( \frac {1+1/\sigma^2 - \frac{1}{v}}  {v(1+1/\sigma^2)}\right) \\
\d4& = &\d4 \delta \left( -\frac{\sigma^2}{1+\sigma^2} \cdot \frac{1}{v^2} +\frac{1}{v} \right) \\
\d4& = &\d4 \delta \left( -\frac{\sigma^2}{1+\sigma^2} \left(\frac{1}{v} - \frac{1+\sigma^2}{2 \sigma^2} \right)^2  + \frac{1+\sigma^2}{4\sigma^2} \right) \\
\d4& \leq &\d4  \delta \left( \frac{1+\sigma^2}{4\sigma^2} \right), \label{philessbound}
\end{eqnarray}
\ES
which completes the proof.
\end{IEEEproof}
\subsection{Proof of Theorem \ref{Theom4}} \label{rate:GAMP}
First, we prove that $R_{\text{AC}} \to I(y;x)$ as $\delta, R_{\text{C}} \to 0$ with $R_{\text{C}}/\delta$ fixed for the BPSK case as follows.

Recall $\rho = \varphi(v)$ in \eqref{GAMPphi} as
\BE \label{phfbound}
\rho = \varphi(v) = \frac {\delta} {v} \lp 1- \frac  {\mathrm{mmse}(x|\hat{p},y)} {v} \rp \leq \frac {\delta} {v}.
\EE
From Proposition \ref{propsition1}, we can see that $\rho \to 0$ as $\delta \to 0$. Then, we apply Taylor expansion to $\text{mmse}(\mathcal{B},\rho)$ at $\rho = 0$ as \cite{Guo2011}
\BE \label{MMSE_BPSK}
v = \text{mmse} \left( \mathcal{B}, \rho  \right) = 1 - \rho  + o\left( \rho^2  \right).
\EE
Denote $\rho_f$ as solutions of $\varphi^{-1}(\rho) = \text{mmse} \left( \mathcal{B}, \rho  \right)$. Combining \eqref{phfbound} and \eqref{MMSE_BPSK}, we have $\rho_f$ at the vicinity of $\rho = 0$
\BE
\rho_f \leq \frac {\delta} {1-\rho_f} \overset{(a)}{\Longrightarrow} \left( \rho_f - \frac {1} {2} \right)^2 \leq  \frac {1} {4} -  \delta \overset{(b)}{\Longrightarrow} 0 \leq \rho_f \leq   \frac {1} {2} - \sqrt{\frac {1} {4} -  \delta}, \label{rhofbound}
\EE
where ($a$) and ($b$) are due to that $\delta \to 0$ and $\rho \to 0$. Define $\rho_f^{max} \triangleq \frac {1} {2} - \sqrt{\frac {1} {4} -  \delta}$. From \eqref{areaf}, we have
\BE \label{arearectangle}
a_f \le \int_0^{\rho_f} {\left( 1 - \text{mmse}  (\mathcal{B}, \rho)  \right){\rm{d}}} \rho \le \int_0^{\rho_f^{max}} {\left( 1 - \text{mmse}  (\mathcal{B}, \rho)  \right){\rm{d}}} \rho \triangleq a_f^{max}.
\EE
Following \eqref{rhoB} and \eqref{areaBPSK}--\eqref{Eqn:Delta_asym}, we can show $\frac {(a_f^{max})^2} {2\delta} \to 0$ as $\delta \to 0$. Since $a_f \leq a_f^{max}$, we have $a_{f}/ 2\delta \rightarrow 0$ as $\delta \to 0$, which completes the proof for the BPSK case.

Following the same method in the proof of Theorem \ref{Theom2}, the above result can be extended to a general symmetrical constellation $\mathcal{S}_C$.
\vspace{-0.5cm}\subsection{Extension of Theorem \ref{Theom4}} \label{theo4extension} \vspace{-0.4cm}
The proof of Theorem \ref{Theom4} relies on Proposition \ref{propsition1} to show that $\rho$ is in the vicinity of zero when $\delta \to 0$. In fact, Theorem \ref{Theom4} holds for an arbitrary signal model as long as Proposition \ref{propsition1} holds. Another provable example is $\iby = f(\ibx+\ibn)$. When $f$ is a one-to-one mapping function, it can be shown that $\textrm{mmse}(x|\hat{p},y) = \frac {v \sigma^2} {v+\sigma^2}$, which is the MMSE for $\iby = \ibx+\ibn$. When $f$ is a many-to-one mapping function, we have $\textrm{mmse}(x|\hat{p},y) > \frac {v \sigma^2} {v+\sigma^2}$ due to the ambiguity from $\iby$ to $\ibx+\ibn$. Thus, we have $\textrm{mmse}(x|\hat{p},y) \geq \frac {v \sigma^2} {v+\sigma^2}$. Combining with \eqref{GAMPphi2} yields
\BE
\rho = \varphi(v) \leq \delta \left( v+ \frac {1} {\frac {v+\sigma^2}{v\sigma^2} - \frac{1}{v} } \right)^{-1} \leq \frac {\delta} {v+\sigma^2} \leq \frac {\delta} {\sigma^2}.
\EE
It is clear that $\rho \to 0$ as $\delta \to 0$.
\end{appendices}\vspace{-0.5cm}
\bibliographystyle{IEEEtran}
\bibliography{MI_GAMP}

\begin{thebibliography}{10}
\providecommand{\url}[1]{#1}
\csname url@samestyle\endcsname
\providecommand{\newblock}{\relax}
\providecommand{\bibinfo}[2]{#2}
\providecommand{\BIBentrySTDinterwordspacing}{\spaceskip=0pt\relax}
\providecommand{\BIBentryALTinterwordstretchfactor}{4}
\providecommand{\BIBentryALTinterwordspacing}{\spaceskip=\fontdimen2\font plus
\BIBentryALTinterwordstretchfactor\fontdimen3\font minus
  \fontdimen4\font\relax}
\providecommand{\BIBforeignlanguage}[2]{{%
\expandafter\ifx\csname l@#1\endcsname\relax
\typeout{** WARNING: IEEEtran.bst: No hyphenation pattern has been}%
\typeout{** loaded for the language `#1'. Using the pattern for}%
\typeout{** the default language instead.}%
\else
\language=\csname l@#1\endcsname
\fi
#2}}
\providecommand{\BIBdecl}{\relax}
\BIBdecl

\bibitem{Liang16}
C.~Liang, J.~Ma, and L.~Ping, ``Towards {Gaussian} capacity, universality and
  short block length,'' \emph{Proc. 9th Int. Symp. Turbo Codes (ISTC)}, pp.
  412--416, Sep 2016.

\bibitem{Liang2020}
S.~{Liang}, C.~{Liang}, J.~{Ma}, and L.~{Ping}, ``Compressed coding and analog
  spatial coupling using {AMP} based decoding,'' \emph{submitted to 2020 IEEE
  Global Communications Conference (GLOBECOM)}, 2020.

\bibitem{cLiang17}
C.~{Liang}, J.~{Ma}, and L.~{Ping}, ``Compressed {FEC} codes with
  spatial-coupling,'' \emph{IEEE Commun. Lett.}, vol.~21, no.~5, pp. 987--990,
  May 2017.

\bibitem{Ma19}
J.~{Ma}, L.~{Liu}, X.~{Yuan}, and L.~{Ping}, ``On orthogonal {AMP} in coded
  linear vector systems,'' \emph{IEEE Transactions on Wireless Communications},
  vol.~18, no.~12, pp. 5658--5672, Dec 2019.

\bibitem{liu2019capacity}
L.~Liu, C.~Liang, J.~Ma, and L.~Ping, ``Capacity optimality of {AMP} in coded
  systems,'' \emph{arXiv:1901.09559}, 2019.

\bibitem{Barron12}
A.~R. Barron and A.~Joseph, ``Least squares superposition codes of moderate
  dictionary size are reliable at rates up to capacity,'' \emph{IEEE Trans.
  Inf. Theory}, vol.~58, no.~2, pp. 2541--2557, Feb. 2012.

\bibitem{Joseph14}
A.~Joseph and A.~R. Barron, ``Fast sparse superposition codes have near
  exponential error probability for ${R} < \cal {C}$,'' \emph{IEEE Trans. Inf.
  Theory}, vol.~60, no.~2, pp. 919--942, Feb. 2014.

\bibitem{SRC19}
R.~Venkataramanan, S.~Tatikonda, and A.~Barron, ``Sparse regression codes,''
  \emph{Foundations and Trends in Communications and Information Theory},
  vol.~15, no. 1-2, pp. 1--195, 2019.

\bibitem{Donoho2009}
D.~L. Donoho, A.~Maleki, and A.~Montanari, ``Message-passing algorithms for
  compressed sensing,'' in \emph{Proc. Nat. Acad. Sci.}, vol. 106, no.~45, Nov.
  2009, pp. 18\,914--18\,919.

\bibitem{Rush17}
C.~{Rush}, A.~{Greig}, and R.~{Venkataramanan}, ``Capacity-achieving sparse
  superposition codes via approximate message passing decoding,'' \emph{IEEE
  Trans. Inf. Theory}, vol.~63, no.~3, pp. 1476--1500, Mar 2017.

\bibitem{Barbier17AMP-decoder}
J.~{Barbier} and F.~{Krzakala}, ``Approximate message-passing decoder and
  capacity achieving sparse superposition codes,'' \emph{IEEE Trans. Inf.
  Theory}, vol.~63, no.~8, pp. 4894--4927, Aug 2017.

\bibitem{SEBayati}
M.~Bayati and A.~Montanari, ``The dynamics of message passing on dense graphs,
  with applications to compressed sensing,'' \emph{IEEE Trans. Inf. Theory},
  vol.~57, no.~2, pp. 764--785, Feb 2011.

\bibitem{Javanmard13}
A.~Javanmard and A.~Montanari, ``State evolution for general approximate
  message passing algorithms, with applications to spatial coupling,''
  \emph{Inf. Inference}, vol.~2, no.~2, pp. 115--144, 2013.

\bibitem{Jeon15}
C.~{Jeon}, R.~{Ghods}, A.~{Maleki}, and C.~{Studer}, ``Optimality of large
  {MIMO} detection via approximate message passing,'' in \emph{2015 IEEE
  International Symposium on Information Theory (ISIT)}, 2015, pp. 1227--1231.

\bibitem{Kudekar13}
S.~Kudekar, T.~Richardson, and R.~L. Urbanke, ``Spatially coupled ensembles
  universally achieve capacity under belief propagation,'' \emph{IEEE Trans.
  Inf. Theory}, vol.~59, no.~12, pp. 7761--7813, Dec. 2013.

\bibitem{Mitchell15}
D.~G.~M. Mitchell, M.~Lentmaier, and D.~J. Costello, Jr., ``Spatially coupled
  {LDPC} codes constructed from protographs,'' \emph{IEEE Trans. Inf. Theory},
  vol.~61, no.~9, pp. 4866--4889, Sep. 2015.

\bibitem{Ma15}
X.~Ma, C.~Liang, K.~Huang, and Q.~Zhuang, ``Block {M}arkov superposition
  transmission: {C}onstruction of big convolutional codes from short codes,''
  \emph{IEEE Trans. Inf. Theory}, vol.~61, no.~6, pp. 3150--3163, Jun. 2015.

\bibitem{Hou16}
W.~Hou, S.~Lu, and J.~Cheng, ``Spatially coupled
  repeater-combiner-convolutional codes,'' \emph{IEEE Commun. Lett.}, vol.~20,
  no.~1, pp. 21--24, Jan. 2016.

\bibitem{Takeuchi11}
K.~Takeuchi, T.~Tanaka, and T.~Kawabata, ``Improvement of {BP}-based {CDMA}
  multiuser detection by spatial coupling,'' in \emph{Proc. IEEE Int. Symp.
  Inf. Theory}, St. Petersburg, Russian, Jul 2011, pp. 1489--1493.

\bibitem{Truhachev13a}
D.~{Truhachev} and C.~{Schlegel}, ``Coupling data transmission for
  multiple-access communications,'' \emph{IEEE Trans. Inf. Theory}, vol.~65,
  no.~7, pp. 4550--4574, Jul 2019.

\bibitem{Kudekar10}
S.~{Kudekar} and H.~D. {Pfister}, ``The effect of spatial coupling on
  compressive sensing,'' in \emph{2010 48th Annual Allerton Conference on
  Communication, Control, and Computing (Allerton)}, Sep. 2010, pp. 347--353.

\bibitem{Krzakala12}
F.~Krzakala, M.~M{\'e}zard, F.~Sausset, Y.~Sun, and L.~Zdeborov{\'a},
  ``{Statistical physics-based reconstruction in compressed sensing},''
  \emph{{Physical Review X}}, vol.~2, p. 021005, 2012.

\bibitem{Donoho2012}
D.~L. {Donoho}, A.~{Javanmard}, and A.~{Montanari}, ``Information-theoretically
  optimal compressed sensing via spatial coupling and approximate message
  passing,'' \emph{IEEE Trans. Inf. Theory}, vol.~59, no.~11, pp. 7434--7464,
  Nov 2013.

\bibitem{Barbier17GAMP}
J.~{Barbier}, M.~{Dia}, and N.~{Macris}, ``Universal sparse superposition codes
  with spatial coupling and {GAMP} decoding,'' \emph{IEEE Trans. Inf. Theory},
  vol.~65, no.~9, pp. 5618--5642, Sep. 2019.

\bibitem{Rush20}
C.~Rush, K.~Hsieh, and R.~Venkataramanan, ``Capacity-achieving spatially
  coupled sparse superposition codes with {AMP} decoding,''
  \emph{arXiv:2002.07844}, 2020.

\bibitem{Hsieh20}
K.~Hsieh and R.~Venkataramanan, ``Modulated sparse superposition codes for the
  complex {AWGN} channel,'' \emph{arXiv:2004.09549}, 2020.

\bibitem{sLiang17}
S.~{Liang}, J.~{Ma}, and L.~{Ping}, ``Clipping can improve the performance of
  spatially coupled sparse superposition codes,'' \emph{IEEE Commun. Lett.},
  vol.~21, no.~12, pp. 2578--2581, Dec 2017.

\bibitem{Yue07}
G.~{Yue}, L.~{Ping}, and X.~{Wang}, ``Generalized low-density parity-check
  codes based on {Hadamard} constraints,'' \emph{IEEE Trans. Inf. Theory},
  vol.~53, no.~3, pp. 1058--1079, Mar 2007.

\bibitem{Divsalar11}
S.~{Abu-Surra}, D.~{Divsalar}, and W.~E. {Ryan}, ``Enumerators for
  protograph-based ensembles of {LDPC} and generalized {LDPC} codes,''
  \emph{IEEE Trans. Inf. Theory}, vol.~57, no.~2, pp. 858--886, Feb 2011.

\bibitem{Ping03}
{Li Ping}, W.~K. {Leung}, and K.~Y. {Wu}, ``Low-rate {turbo-Hadamard} codes,''
  \emph{IEEE Trans. Inf. Theory}, vol.~49, no.~12, pp. 3213--3224, Dec 2003.

\bibitem{Linshu04}
S.~Lin and D.~J. Costello, \emph{Error control coding}.\hskip 1em plus 0.5em
  minus 0.4em\relax Pearson Education India, 2004.

\bibitem{Huang07}
Q.~{Huang}, S.~{Chan}, L.~{Ping}, and M.~{Zukerman}, ``Improving wireless {TCP}
  throughput by a novel {TCM-Based} hybrid {ARQ},'' \emph{IEEE Trans. Wirel.
  Commun.}, vol.~6, no.~7, pp. 2476--2485, Jul 2007.

\bibitem{Zhang09}
R.~{Zhang} and L.~{Hanzo}, ``Superposition-coding-aided multiplexed hybrid
  {ARQ} scheme for improved end-to-end transmission efficiency,'' \emph{IEEE
  Transactions on Vehicular Technology}, vol.~58, no.~8, pp. 4681--4686, Oct
  2009.

\bibitem{Leung06}
W.~K.~R. Leung, G.~Yue, L.~Ping, and X.~Wang, ``Concatenated zigzag {H}adamard
  codes,'' \emph{IEEE Trans. Inf. Theory}, vol.~52, no.~4, pp. 1711--1723, Apr.
  2006.

\bibitem{Ping10}
L.~{Ping}, J.~{Tong}, X.~{Yuan}, and Q.~{Guo}, ``Superposition coded modulation
  and iterative linear {MMSE} detection,'' \emph{IEEE Journal on Selected Areas
  in Communications}, vol.~27, no.~6, pp. 995--1004, 2009.

\bibitem{Bhattad2007}
K.~Bhattad and K.~Narayanan, ``An {MSE}-based transfer chart for analyzing
  iterative decoding schemes using a {G}aussian approximation,'' \emph{IEEE
  Trans. Inf. Theory}, vol.~53, no.~1, pp. 22--38, Jan. 2007.

\bibitem{Guo2011}
D.~Guo, Y.~Wu, S.~Shamai, and S.~Verdu, ``Estimation in {Gaussian} noise:
  properties of the minimum mean-square error,'' \emph{IEEE Trans. Inf.
  Theory}, vol.~57, no.~4, pp. 2371--2385, Apr 2011.

\bibitem{GAMP}
S.~Rangan, ``Generalized approximate message passing for estimation with random
  linear mixing,'' in \emph{IEEE International Symposium on Information Theory
  (ISIT)}, Jul 2011, pp. 2168--2172.

\bibitem{Kudekar15}
S.~{Kudekar}, T.~J. {Richardson}, and R.~L. {Urbanke}, ``Wave-like solutions of
  general {1-D} spatially coupled systems,'' \emph{IEEE Trans. Inf. Theory},
  vol.~61, no.~8, pp. 4117--4157, Aug 2015.

\bibitem{Biyik17}
E.~{Biyik}, J.~{Barbier}, and M.~{Dia}, ``Generalized approximate
  message-passing decoder for universal sparse superposition codes,'' in
  \emph{IEEE Int. Symp. Inf. Theory (ISIT)}, Jun 2017, pp. 1593--1597.

\bibitem{Yedla14}
A.~Yedla, Y.-Y. Jian, P.~S. Nguyen, and H.~D. Pfister, ``A simple proof of
  {M}axwell saturation for coupled scalar recursions,'' \emph{IEEE Trans. Inf.
  Theory}, vol.~60, no.~11, pp. 6943--6965, Nov 2014.

\bibitem{Borst10}
S.~{Borst}, V.~{Gupta}, and A.~{Walid}, ``Distributed caching algorithms for
  content distribution networks,'' in \emph{2010 Proceedings IEEE INFOCOM},
  March 2010, pp. 1--9.

\bibitem{Ping06}
L.~Ping, L.~Liu, K.~Wu, and W.~K. Leung, ``Interleave division
  multiple-access,'' \emph{IEEE Trans. Wirel. Commun.}, vol.~5, no.~4, pp.
  938--947, Apr. 2006.

\bibitem{Tal15}
I.~Tal and A.~Vardy, ``List decoding of polar codes,'' \emph{IEEE Trans. Inf.
  Theory}, vol.~61, no.~5, pp. 2213--2226, May 2015.

\bibitem{Yedla11}
A.~Yedla, P.~S. Nguyen, H.~D. Pfister, and K.~R. Narayanan, ``Universal codes
  for the {Gaussian} {MAC} via spatial coupling,'' in \emph{{Proc. Allerton
  Conf. Commun., Contr. \& Comput.}}, Monticello, IL, USA, Sep 2011, pp.
  1801--1808.

\bibitem{Ma04}
X.~Ma and L.~Ping, ``Coded modulation using superimposed binary codes,''
  \emph{IEEE Trans. Inf. Theory}, vol.~50, no.~12, pp. 3331--3343, Dec. 2004.

\end{thebibliography}
\end{document}